\documentclass[preprint,superscriptaddress]{revtex4-1} 
\usepackage{graphicx} 
\usepackage{amsmath} 
\usepackage[version=3]{mhchem} 
\usepackage{longtable} 
\usepackage{filecontents} 
\usepackage{cleveref} 
\usepackage{multirow} 
\usepackage{mathtools} 
\usepackage{url} 
\usepackage{verbatim} 
 
\usepackage{epstopdf} 
\usepackage{soul} 
\usepackage[english]{babel} 
\usepackage{chemfig} 

\usepackage{subfigure} 
\usepackage{amsmath}
\usepackage{amssymb}

\usepackage{physics}

\usepackage{tabularray}
\usepackage{xr}

\begin{document} 

\title{Finite-Temperature Thermally-Assisted-Occupation Density Functional Theory, \textit{Ab Initio} Molecular Dynamics, and Quantum Mechanics/Molecular Mechanics Methods} 

\author{Shaozhi Li} 
\affiliation{Department of Physics, National Taiwan University, Taipei 10617, Taiwan} 

\author{Jeng-Da Chai} 
\email[Author to whom correspondence should be addressed. Electronic mail: ]{jdchai@phys.ntu.edu.tw} 
\affiliation{Department of Physics, National Taiwan University, Taipei 10617, Taiwan} 
\affiliation{Center for Theoretical Physics and Center for Quantum Science and Engineering, National Taiwan University, Taipei 10617, Taiwan} 
\affiliation{Physics Division, National Center for Theoretical Sciences, Taipei 10617, Taiwan} 

\date{\today} 

\begin{abstract} 

Recently, thermally-assisted-occupation density functional theory (TAO-DFT) [J.-D. Chai, J. Chem. Phys. {\bf 136}, 154104 (2012)] has been demonstrated to be an efficient and accurate electronic structure method for 
studying the ground-state properties of large multi-reference (MR) systems at absolute zero. To explore the thermal equilibrium properties of large MR systems at finite electronic temperatures, in the present work, we 
propose the finite-temperature (FT) extension of TAO-DFT, denoted as FT-TAO-DFT. Besides, to unlock the dynamical information of large MR systems at finite temperatures, FT-TAO-DFT is combined with \textit{ab initio} 
molecular dynamics, leading to FT-TAO-AIMD. In addition, we also develop FT-TAO-DFT-based quantum mechanics/molecular mechanics (QM/MM), denoted as FT-TAO-QM/MM, to provide a cost-effective description of 
the thermal equilibrium properties of a QM subsystem with MR character embedded in an MM environment at finite temperatures. Moreover, the FT-TAO-DFT, FT-TAO-AIMD, and FT-TAO-QM/MM methods are employed 
to explore the radical nature and infrared (IR) spectra of $n$-acenes ($n$ = 2--6), consisting of $n$ linearly fused benzene rings, in vacuum and in an argon (Ar) matrix at finite temperatures. According to our calculations, 
for $n$-acenes at 1000 K or below, the electronic temperature effects on the radical nature and IR spectra are very minor, while the nuclear temperature effects on these properties are noticeable. For $n$-acene in an Ar 
matrx at absolute zero, the Ar matrix has minimal impact on the radical nature of $n$-acene, while the co-deposition procedure of $n$-acene and Ar atoms may affect the IR spectrum of $n$-acene. 

\end{abstract} 

\maketitle

\section{Introduction} 

In the last three decades, Kohn-Sham density functional theory (KS-DFT) \cite{HK,KS-DFT} has become a popular electronic structure method for exploring the ground-state (GS) properties of large electronic systems at 
absolute zero, due to its proper balance between efficiency and accuracy \cite{parr-yang,Jensen,Kohanoff,DFTReview,Yang12}. The finite-temperature (FT) extension of KS-DFT (FT-KS-DFT) \cite{Mermin,KS-DFT}, 
commonly known as finite-temperature density functional theory (FT-DFT) or the Mermin-Kohn-Sham (MKS) method, has also been developed to study the thermal equilibrium (TE) properties of large electronic systems 
at finite electronic temperatures ($\theta_{el} \equiv k_{B} T_{el} \ge 0$, where $\theta_{el}$ is the electronic temperature measured in energy units, $T_{el}$ is the electronic temperature measured in absolute temperature, 
and $k_{B}$ is the Boltzmann constant) \cite{gross-2011}. 

For the GS properties of electronic systems at zero electronic temperature ($\theta_{el} = 0$), FT-KS-DFT with the exchange-correlation (xc) free energy functional reduces to KS-DFT with the xc energy functional. Note, 
however, that KS-DFT with the conventional local density approximation (LDA) \cite{LDAX,LDAC}, generalized gradient approximation (GGA) \cite{B88,LYP,PBE}, global hybrid (GH) \cite{hybrid1,hybrid2,hybrid0}, and 
range-separated hybrid (RSH) \cite{LC-DFT,LCHirao,wB97X} xc energy functionals can perform poorly for the GS properties of multi-reference (MR) systems (i.e., systems where the electronic GS wavefunctions are not 
dominated by single Slater determinants) at $\theta_{el} = 0$, due to the lack of proper treatment of the GS density representability and static correlation energy \cite{Yang-science-2008,Yang-JCP-2008,Yang12}. In particular, 
for a singlet GS system with pronounced MR character, KS-DFT with the conventional xc energy functionals can yield incorrect spin orbitals, spin densities, and related properties, wherein the spin-unrestricted properties can 
differ greatly from the spin-restricted properties, yielding the unphysical spin-symmetry breaking effects in the spin-unrestricted properties. 

At very low electronic temperatures ($\theta_{el} \approx 0$), the electronic thermal ensembles of an MR system are mainly contributed by the electronic GS (i.e., the electronic thermal ensemble at $\theta_{el} = 0$). 
Therefore, at $\theta_{el} \approx 0$, the TE properties of MR systems are largely dominated by the corresponding GS properties at $\theta_{el} = 0$. Accordingly, it can be anticipated that FT-KS-DFT with the conventional 
LDA \cite{ftlda}, GGA \cite{ftgga}, GH \cite{ftgh}, and RSH \cite{ftrsh} xc free energy functionals can also perform poorly for the TE properties of MR systems at $\theta_{el} \approx 0$. 

In general, \textit{ab initio} MR electronic structure methods \cite{CASSCF,CASPT2,dmrg,2-RDMa,cote2015} are needed to reliably predict the properties of MR systems at absolute zero. Nonetheless, owing to their 
prohibitively high computational complexity, it remains infeasible to perform accurate MR electronic structure calculations for large MR systems at zero electronic temperature ($\theta_{el} = 0$), not to mention the respective 
calculations at finite electronic temperatures ($\theta_{el} \ge 0$). 

Aiming to improve the GS density representability and static correlation energy for MR systems at zero electronic temperature ($\theta_{el} = 0$), thermally-assisted-occupation density functional theory (TAO-DFT) \cite{tao1} 
has been recently developed. In contrast to KS-DFT, TAO-DFT allows fractional orbital occupations (described by the Fermi-Dirac (FD) distribution with some fictitious temperature $\theta$ (i.e., the temperature of the 
non-interacting reference systems in TAO-DFT)). To improve the GS density representability \cite{tao1}, the fictitious temperature $\theta$ in TAO-DFT can be so selected that the orbitals and their occupation numbers 
approximately describe the exact natural orbitals (NOs) and natural orbital occupation numbers (NOONs) \cite{Lowdin-1955,Lowdin-1956}, respectively. Therefore, the fictitious temperature $\theta$ should be very small for 
single-reference (SR) systems (i.e., systems where the electronic GS wavefunctions are dominated by single Slater determinants), and highly system-dependent for MR systems \cite{tao1}. Recently, it has been proved that 
TAO-DFT with a sufficiently large $\theta$ can always resolve the aforementioned unphysical spin-symmetry breaking problems for singlet GS systems (i.e., very challenging problems for KS-DFT with the conventional xc 
energy functionals) \cite{spin-symm}, highlighting the significance of TAO-DFT. Note also that TAO-DFT (with a fictitious temperature $\theta$) is as computationally efficient as KS-DFT (i.e., TAO-DFT with $\theta = 0$). 

Since the exact exchange-correlation-$\theta$ (xc$\theta$) energy functional (i.e., a combined xc and $\theta$-dependent energy functional) \cite{tao1}, which is the essential ingredient of TAO-DFT, remains unknown, an 
approximate xc$\theta$ energy functional has to be employed for practical TAO-DFT calculations. The LDA \cite{tao1}, GGA \cite{tao2}, GH \cite{tao3}, and RSH \cite{tao3,tao-rsh} xc$\theta$ energy functionals in TAO-DFT 
have been developed in recent years. For a given xc$\theta$ energy functional in TAO-DFT, the optimal system-independent \cite{tao1,tao2,tao3,theta2022} and system-dependent \cite{tao4} $\theta$-schemes have been 
recently proposed to determine the optimal $\theta$. With an appropriately selected $\theta$, the static correlation energy of an electronic system can be approximately described by the entropy contribution in 
TAO-DFT \cite{tao1,tao2,tao3}. Very recently, we have assessed the performance of various KS-DFT and TAO-DFT functionals on a very wide range of test sets \cite{tao-rsh}, including both SR and MR systems. According 
to our study, the KS-DFT functionals can perform very poorly for MR systems, while the TAO-DFT functionals can achieve reasonably good performance for both SR and MR systems. 

Because of its reasonable accuracy and computational efficiency, TAO-DFT has been widely applied to study the GS properties of various MR systems at the nanoscale \cite{tao-gnr,NK,H2S1,cycl,coronene,mobius,BNR,CC,
SS21,CCC,SS24,Efields,Manassir2020,HansonHeine2020,HansonHeine2020b,HansonHeine2020c,Bettinger2021,Bettinger2021a,tao-vib2,Nieman2023,Bettinger2024,Bettinger2025,HansonHeine2025}. Recently, several 
extensions of TAO-DFT have also been proposed, including TAO-DFT-based \textit{ab initio} molecular dynamics (TAO-AIMD) \cite{tao-aimd}, TAO-DFT with the polarizable continuum model (TAO-PCM) \cite{TAO-PCM}, 
a real-time extension of TAO-DFT (RT-TAO-DFT) \cite{rttao}, and a TAO-DFT-based excited-state method (pTAO/TDA) \cite{tao-rsh}. 

Similar to KS-DFT, TAO-DFT is a GS electronic structure method. Aiming to explore the TE properties of large MR systems at finite electronic temperatures ($\theta_{el} \ge 0$), in this work, we propose the FT extension of 
TAO-DFT, denoted as FT-TAO-DFT. Besides, to unlock the dynamical information of large MR systems at finite temperatures, we combine FT-TAO-DFT with \textit{ab initio} molecular dynamics (AIMD) \cite{Jensen,AIMD2002}, 
yielding FT-TAO-AIMD. In addition, we also develop FT-TAO-DFT-based quantum mechanics/molecular mechanics (QM/MM) \cite{QMMM-1,QMMM-2,QMMM-3}, denoted as FT-TAO-QM/MM, to provide a cost-effective 
description of the TE properties of a QM subsystem with MR character embedded in an MM environment at finite temperatures. To demonstrate some of their capacities, the FT-TAO-DFT, FT-TAO-AIMD, and FT-TAO-QM/MM 
methods are employed to explore the radical nature and infrared (IR) spectra of $n$-acenes ($n$ = 2--6), composed of $n$ linearly fused benzene rings, in vacuum and in an argon (Ar) matrix at finite temperatures. 

The rest of this paper is organized as follows. The FT-TAO-DFT, FT-TAO-AIMD, and FT-TAO-QM/MM methods are proposed in \Cref{sec:fttao}, \Cref{sec:ft-aimd}, and \Cref{tao-qmmm}, respectively. The computational details 
are described in \Cref{sec:comp}. The radical nature and IR spectra of $n$-acenes ($n$ = 2--6) in vacuum and in an Ar matrix at finite temperatures are presented and discussed in \Cref{sec:results}. Our conclusions are given 
in \Cref{conclusion}.

\section{FT-TAO-DFT}\label{sec:fttao} 

\subsection{Spin-polarized formalism} 

Consider the electronic thermal ensemble of a physical system which on average, contains $N_{\alpha}$ $\alpha$-spin and $N_{\beta}$ $\beta$-spin interacting electrons in an external potential $v_{\text{ext}}({\bf r})$ at the 
electronic temperature $\theta_{el} \equiv k_{B} T_{el} \ge 0$. In spin-polarized (spin-unrestricted) FT-TAO-DFT, the $\sigma$-spin TE density $\rho_{\sigma}({\bf r})$ (with $\sigma$ = $\alpha$ or $\beta$) of the physical 
system is represented by the TE density $\rho_{s,\sigma}({\bf r})$ of a non-interacting reference system at some fictitious temperature $\theta$. The two non-interacting reference systems (i.e., one described by the spin 
function $\alpha$ and the other described by the spin function $\beta$) are hereafter referred to as the FT-TAO reference systems. 

Following the similar derivations of the self-consistent equations in spin-polarized TAO-DFT (e.g., see Section III.A and III.B of the TAO-DFT paper \cite{tao1}) for the GS density of a physical system at zero electronic 
temperature ($\theta_{el}$ = 0), the self-consistent equations in spin-polarized FT-TAO-DFT for the TE density of a physical system at the electronic temperature ($\theta_{el} \ge 0$) can be straightforwardly obtained, based 
on the Mermin theorems \cite{Mermin,Mermin-SP-0,Mermin-SP-1,Mermin-SP-2} for both the physical system at the electronic temperature $\theta_{el}$ and the FT-TAO reference systems at the fictitious temperature $\theta$. 

In spin-polarized FT-TAO-DFT, to obtain the $\sigma$-spin TE density $\rho_{\sigma}({\bf r})$, the self-consistent equations (atomic units (a.u.) are adopted in this work unless otherwise noted) are given 
by ($i$ runs for the orbital index) 
\begin{equation}\label{eq:tao-eq1} 
\bigg\lbrace - \frac{1}{2} {\bf \nabla}^{2} \ + \ v_{s,\sigma}({\bf r}) \bigg\rbrace \psi_{i\sigma}({\bf r}) = \epsilon_{i\sigma} \psi_{i\sigma}({\bf r}), 
\end{equation} 
with 
\begin{equation}\label{eq:tao-eq2} 
v_{s,\sigma}({\bf r}) = v_{\text{ext}}({\bf r}) + \frac{\delta E_{\text{H}}[\rho]}{\delta \rho({\bf r})} + \frac{\delta F_{\text{xc}\theta}^{\theta_{el}}[\rho_{\alpha},\rho_{\beta}]}{\delta \rho_{\sigma}({\bf r})} 
\end{equation} 
being the $\sigma$-spin effective one-electron potential. In Eq.\ (\ref{eq:tao-eq2}), 
\begin{equation} 
E_{\text{H}}[\rho] \equiv \frac{1}{2} \iint \frac{\rho({\bf r})\rho({\bf r'})}{|{\bf r} - {\bf r'}|}d{\bf r}d{\bf r'} 
\end{equation} 
is the Hartree energy functional, and 
\begin{equation}\label{eq:ftxcheta} 
F_{\text{xc}\theta}^{\theta_{el}}[\rho_{\alpha},\rho_{\beta}] \equiv F_{\text{M}}^{\theta_{el}}[\rho_{\alpha},\rho_{\beta}] - A_{s}^{\theta}[\rho_{\alpha},\rho_{\beta}] - E_{\text{H}}[\rho] 
\end{equation} 
is the xc$\theta$ free energy functional at temperature $\theta_{el}$, where $F_{\text{M}}^{\theta_{el}}[\rho_{\alpha},\rho_{\beta}]$ is the Mermin universal 
functional \cite{Mermin,gross-2011,Mermin-SP-0,Mermin-SP-1,Mermin-SP-2} (i.e., the sum of the interacting kinetic free energy and the electron-electron repulsion energy) at temperature $\theta_{el}$, and 
$A_{s}^{\theta}[\rho_{\alpha},\rho_{\beta}]$ is the non-interacting kinetic free energy functional at temperature $\theta$. 
The xc$\theta$ free energy functional at temperature $\theta_{el}$, $F_{\text{xc}\theta}^{\theta_{el}}[\rho_{\alpha},\rho_{\beta}]$ (defined by Eq.\ (\ref{eq:ftxcheta})), can be usefully partitioned into the following terms: 
\begin{equation}\label{eq:ftxcheta2} 
F_{\text{xc}\theta}^{\theta_{el}}[\rho_{\alpha},\rho_{\beta}] = F_{\text{xc}}^{\theta_{el}}[\rho_{\alpha},\rho_{\beta}] + F_{\theta}^{\theta_{el}}[\rho_{\alpha},\rho_{\beta}], 
\end{equation} 
where 
\begin{equation} 
F_{\text{xc}}^{\theta_{el}}[\rho_{\alpha},\rho_{\beta}] \equiv F_{\text{M}}^{\theta_{el}}[\rho_{\alpha},\rho_{\beta}] - A_{s}^{\theta_{el}}[\rho_{\alpha},\rho_{\beta}] - E_{\text{H}}[\rho] 
\end{equation} 
is the xc free energy functional at temperature $\theta_{el}$ (as defined in FT-KS-DFT \cite{KS-DFT,Mermin,gross-2011,Mermin-SP-0,Mermin-SP-1,Mermin-SP-2}), and 
\begin{equation}\label{eq:ftheta} 
F_{\theta}^{\theta_{el}}[\rho_{\alpha},\rho_{\beta}] \equiv A_{s}^{\theta_{el}}[\rho_{\alpha},\rho_{\beta}] - A_{s}^{\theta}[\rho_{\alpha},\rho_{\beta}], 
\end{equation} 
is the $\theta$-dependent free energy functional at temperature $\theta_{el}$, with $A_{s}^{\theta_{el}}[\rho_{\alpha},\rho_{\beta}]$ being the non-interacting kinetic free energy functional at temperature $\theta_{el}$. 
By construction, the $\sigma$-spin TE density $\rho_{\sigma}({\bf r})$ is represented by $\rho_{s,\sigma}({\bf r})$, which can be expressed as 
\begin{equation}\label{eq:tao-eq3} 
\rho_{\sigma}({\bf r}) = \rho_{s,\sigma}({\bf r}) = \sum_{i=1}^{\infty} f_{i\sigma} |\psi_{i\sigma}({\bf r})|^{2}. 
\end{equation} 
Here, $f_{i\sigma}$ is the occupation number of the $i$-th $\sigma$-spin orbital $\psi_{i\sigma}({\bf r})$, which is described by the FD distribution 
\begin{equation}\label{eq:tao-eq4} 
f_{i\sigma} = \{1 + \text{exp}[(\epsilon_{i\sigma} - \mu_{\sigma}) / \theta]\}^{-1}, 
\end{equation} 
where $\epsilon_{i\sigma}$ is the energy of $\psi_{i\sigma}({\bf r})$, and $\mu_{\sigma}$ is the $\sigma$-spin chemical potential chosen to conserve $N_{\sigma}$ (i.e., the average number of $\sigma$-spin electrons): 
\begin{equation}\label{eq:tao-eq5} 
\sum_{i=1}^{\infty} \{1 + \text{exp}[(\epsilon_{i\sigma} - \mu_{\sigma}) / \theta]\}^{-1} = N_{\sigma}. 
\end{equation} 
The TE density $\rho({\bf r})$ is computed using 
\begin{equation}\label{eq:tao-eq6} 
\rho({\bf r}) = \sum_{\sigma}^{\alpha,\beta} \rho_{\sigma}({\bf r}). 
\end{equation} 
\Cref{eq:tao-eq1,eq:tao-eq2,eq:tao-eq3,eq:tao-eq4,eq:tao-eq5,eq:tao-eq6} can be employed to determine the $\sigma$-spin orbitals $\{\psi_{i\sigma}({\bf r})\}$, the $\sigma$-spin orbital occupation numbers 
$\{f_{i\sigma}\}$, the $\sigma$-spin TE density $\rho_{\sigma}({\bf r})$, and the TE density $\rho({\bf r})$ in a self-consistent manner (e.g., see Section III.B of the TAO-DFT paper \cite{tao1}). 

After the self-consistency is achieved, the electronic Helmholtz free energy of the physical system at the electronic temperature $\theta_{el}$ is given by 
\begin{equation}\label{eq:tao-energy} 
F_{\text{FT-TAO-DFT}}^{\theta_{el}}[\rho_{\alpha},\rho_{\beta}] = 
\int \rho({\bf r}) v_{\text{ext}}({\bf r})d{\bf r} + A_{s}^{\theta}[\{f_{i\alpha},\psi_{i\alpha}\},\{f_{i\beta},\psi_{i\beta}\}] + E_{\text{H}}[\rho] + F_{\text{xc}\theta}^{\theta_{el}}[\rho_{\alpha},\rho_{\beta}], 
\end{equation} 
where 
\begin{equation} 
A_{s}^{\theta}[\{f_{i\alpha},\psi_{i\alpha}\},\{f_{i\beta},\psi_{i\beta}\}] = T_{s}^{\theta}[\{f_{i\alpha},\psi_{i\alpha}\},\{f_{i\beta},\psi_{i\beta}\}] + E_{S}^{\theta}[\{f_{i\alpha}\},\{f_{i\beta}\}] 
\end{equation} 
is the non-interacting kinetic free energy at temperature $\theta$, which can be exactly computed (in terms of the $\{\psi_{i\sigma}({\bf r})\}$ and $\{f_{i\sigma}\}$) as the sum of the kinetic energy 
\begin{equation} 
T_{s}^{\theta}[\{f_{i\alpha},\psi_{i\alpha}\},\{f_{i\beta},\psi_{i\beta}\}] = -\frac{1}{2} \sum_{\sigma}^{\alpha,\beta} \sum_{i=1}^{\infty} f_{i\sigma} \int \psi_{i\sigma}^{*}({\bf r}){\bf \nabla}^{2}\psi_{i\sigma}({\bf r})d{\bf r} 
\end{equation} 
and entropy contribution 
\begin{equation} 
E_{S}^{\theta}[\{f_{i\alpha} \}, \{f_{i\beta} \}] = \theta \sum_{\sigma}^{\alpha,\beta} \sum_{i=1}^{\infty} \bigg\lbrace f_{i\sigma}\ \text{ln}(f_{i\sigma}) + (1-f_{i\sigma})\ \text{ln}(1-f_{i\sigma}) \bigg\rbrace 
\end{equation} 
of non-interacting electrons at temperature $\theta$. In Eq.\ (\ref{eq:tao-energy}), the sum of the last three terms yields the Mermin universal functional at temperature $\theta_{el}$, 
$F_{\text{M}}^{\theta_{el}}[\rho_{\alpha},\rho_{\beta}]$ (see Eq.\ (\ref{eq:ftxcheta})). Note also that spin-unpolarized (spin-restricted) FT-TAO-DFT can be formulated by imposing the constraints of 
$\psi_{i\alpha}({\bf r})$ = $\psi_{i\beta}({\bf r})$ and $f_{i\alpha}$ = $f_{i\beta}$ to spin-polarized (spin-unrestricted) FT-TAO-DFT. 

At zero electronic temperature ($\theta_{el} = 0$), FT-TAO-DFT with the xc$\theta$ free energy functional reduces to TAO-DFT with the xc$\theta$ energy functional \cite{tao1}, 
which, at $\theta = 0$, further reduces to KS-DFT with the xc energy functional \cite{KS-DFT}. 

On the other hand, if the constraint $\theta = \theta_{el}$ is additionally imposed, FT-TAO-DFT with the xc$\theta$ free energy functional reduces to FT-KS-DFT with the xc free energy 
functional \cite{KS-DFT,Mermin}, which, at $\theta_{el} = 0$, further reduces to KS-DFT with the xc energy functional \cite{KS-DFT}.

\subsection{Local density approximation} 

Since the exact xc$\theta$ free energy functional $F_{\text{xc}\theta}^{\theta_{el}}[\rho_{\alpha},\rho_{\beta}]$ (see Eq.\ (\ref{eq:ftxcheta2})), in terms of the spin densities $\rho_{\alpha}({\bf r})$ and 
$\rho_{\beta}({\bf r})$, has not been known, it is necessary to employ approximate xc$\theta$ free energy functionals for practical FT-TAO-DFT calculations. 

As the LDA is the simplest density functional approximation, in this work, we adopt the LDA xc$\theta$ free energy functional (i.e., the LDA for the xc$\theta$ free energy functional 
$F_{\text{xc}\theta}^{\theta_{el}}[\rho_{\alpha},\rho_{\beta}]$), given by 
\begin{equation}\label{eq:LDAxctheta} 
F_{\text{xc}\theta}^{\text{LDA},\theta_{el}}[\rho_{\alpha},\rho_{\beta}] = F_{\text{xc}}^{\text{LDA},\theta_{el}}[\rho_{\alpha},\rho_{\beta}] + F_{\theta}^{\text{LDA},\theta_{el}}[\rho_{\alpha},\rho_{\beta}], 
\end{equation} 
where the LDA xc free energy functional $F_{\text{xc}}^{\text{LDA},\theta_{el}}[\rho_{\alpha},\rho_{\beta}]$ is available \cite{ftlda}, and the LDA $\theta$-dependent free energy functional 
\begin{equation}\label{eq:thetaF} 
F_{\theta}^{\text{LDA},\theta_{el}}[\rho_{\alpha},\rho_{\beta}] = A_{s}^{\text{LDA},\theta_{el}}[\rho_{\alpha},\rho_{\beta}] - A_{s}^{\text{LDA},\theta}[\rho_{\alpha},\rho_{\beta}] 
\end{equation} 
can be constructed by Eq.\ (\ref{eq:ftheta}) with the LDA non-interacting kinetic free energy functional $A_{s}^{\text{LDA},\theta_{el}}[\rho_{\alpha},\rho_{\beta}]$ \cite{tao1,perrot-1979}. 
FT-TAO-DFT with the LDA xc$\theta$ free energy functional is hereafter referred to as FT-TAO-LDA. 

At zero electronic temperature ($\theta_{el} = 0$), FT-TAO-LDA reduces to TAO-LDA (i.e., TAO-DFT with the LDA xc$\theta$ energy functional) \cite{tao1}. 
On the other hand, if the constraint $\theta = \theta_{el}$ is additionally imposed, FT-TAO-LDA reduces to FT-KS-LDA (i.e., FT-KS-DFT with the LDA xc free energy functional) \cite{ftlda}. 

To go beyond the simplest FT-TAO-LDA, more sophisticated density functional approximations (e.g., GGAs) for the xc free energy functional \cite{ftgga} and the $\theta$-dependent free energy functional (i.e., 
constructed by Eq.\ (\ref{eq:ftheta}) with the non-interacting kinetic free energy functional \cite{Karasiev-2012}) can also be made in FT-TAO-DFT.

\subsection{Fictitious temperature}\label{theta-fttao} 

Consider the TE density $\rho({\bf r})$ of a physical system at the electronic temperature $\theta_{el}$. For the exact FT-TAO-DFT (i.e., FT-TAO-DFT with the exact xc$\theta$ free energy functional 
$F_{\text{xc}\theta}^{\theta_{el}}[\rho_{\alpha},\rho_{\beta}]$), if the TE density representability (see Eq.\ (\ref{eq:tao-eq3})) can be fulfilled for a set of $\theta$ values, the $\sigma$-spin TE density $\rho_{\sigma}({\bf r})$, 
TE density $\rho({\bf r})$, and electronic Helmholtz free energy $F_{\text{FT-TAO-DFT}}^{\theta_{el}}[\rho_{\alpha},\rho_{\beta}]$ cannot vary with the fictitious temperature $\theta$, while the orbitals and their occupation 
numbers in the TE density $\rho({\bf r})$ can vary with $\theta$. 

Similar to the GS counterpart \cite{Lowdin-1955,tao1,Lowdin-1956}, the TE density $\rho({\bf r})$ of a physical system at the electronic temperature $\theta_{el}$ can be expressed by the $\theta_{el}$-dependent NOs 
$\{\chi_{i}({\bf r})\}$ and NOONs $\{n_{i}\}$, based on the exact finite-temperature reduced-density-matrix-functional theory (FT-RDMFT) \cite{ft-rdmt}: 
\begin{equation}\label{eq:FT-RDM-desnity} 
\rho({\bf r}) = \sum_{i=1}^{\infty} n_{i} |\chi_{i}({\bf r})|^{2}. 
\end{equation} 

Therefore, similar to the GS counterpart (e.g., see Section III.E of the TAO-DFT paper \cite{tao1}), the fictitious temperature $\theta$ in the exact FT-TAO-DFT can be properly chosen to improve the TE density 
representability for a physical system at the electronic temperature $\theta_{el}$. Specifically, for the TE density $\rho({\bf r})$ (see \Cref{eq:tao-eq3,eq:tao-eq4,eq:tao-eq5,eq:tao-eq6}) in the exact FT-TAO-DFT, the 
fictitious temperature $\theta$ can be so chosen that the orbitals and their occupation numbers approximately describe the $\theta_{el}$-dependent NOs and NOONs, respectively (see Eq.\ (\ref{eq:FT-RDM-desnity})), 
obtained with the exact FT-RDMFT. Note that the TE density representability (see Eq.\ (\ref{eq:tao-eq3})) is likely to be fulfilled for this $\theta$, due to the similarity of their TE density representations. Consequently, the 
optimal fictitious temperature $\theta$ in the exact FT-TAO-DFT should be both system-dependent and $\theta_{el}$-dependent. The arguments above can also be applied to define the optimal $\theta$ for FT-TAO-DFT 
with the approximate xc$\theta$ free energy functionals (e.g., FT-TAO-LDA). By contrast, for FT-KS-DFT \cite{KS-DFT,Mermin} (i.e., FT-TAO-DFT with $\theta = \theta_{el}$), the constraint $\theta = \theta_{el}$ seems 
unnecessary, and can, in fact, limit the TE density representability. For example, KS-DFT (i.e., FT-KS-DFT at $\theta_{el} = 0$), with the constraint $\theta = \theta_{el} = 0$, has been shown to have serious problems in 
the GS density representability for MR systems \cite{Yang-science-2008,Yang-JCP-2008,Yang12}, which can be properly addressed by TAO-DFT (i.e., FT-TAO-DFT at $\theta_{el} = 0$), with 
$\theta > 0$ \cite{tao1,tao2,tao3,tao-rsh,spin-symm,theta2022}. 

Nevertheless, in FT-TAO-DFT, it remains very challenging to determine the optimal fictitious temperature $\theta$, which should be both system-dependent and $\theta_{el}$-dependent, due to the lack of 
$\theta_{el}$-dependent NOs and NOONs, obtained from the exact FT-RDMFT or other accurate FT electronic structure methods. 

To make progress, in this work, we propose a simple $\theta$-approximation, adopting the optimal system-independent and $\theta_{el}$-independent fictitious temperature $\theta$. Specifically, for a given xc$\theta$ 
free energy functional in FT-TAO-DFT, the optimal system-independent and $\theta_{el}$-independent $\theta$ is defined as the optimal system-independent $\theta$ at zero electronic temperature ($\theta_{el} = 0$). 
Accordingly, the $\theta$ can be determined by one of the optimal system-independent $\theta$-schemes \cite{tao1,tao2,tao3,theta2022} in TAO-DFT (i.e., FT-TAO-DFT at $\theta_{el} = 0$). By construction, this simple 
$\theta$-approximation is expected to work reasonably well at very low electronic temperatures ($\theta_{el} \approx 0$) \cite{tao1,theta2022,tao-rsh}.

\section{FT-TAO-AIMD}\label{sec:ft-aimd} 

Consider a physical system consisting of $N$ nuclei (treated as classical particles) and $N_{el}$ electrons (treated as quantum particles) in thermal equilibrium, wherein the nuclear temperature $T$ is the same as 
the electronic temperature $T_{el}$. It has been recently shown that the nuclei obey the classical nuclear Hamiltonian \cite{free-energy-surface-2} 
\begin{equation}\label{eq:Hnuc} 
H_{nuc}({\bf R}_{1}, ..., {\bf R}_{N}, {\bf P}_{1}, ..., {\bf P}_{N}; T_{el}) = \sum_{A = 1}^{N} \frac{|{\bf P}_{A}|^{2}}{2 M_{A}} + U_{\text{eff}}({\bf R}_{1}, ..., {\bf R}_{N}; T_{el}), 
\end{equation} 
where $M_{A}$, ${\bf R}_{A}$, and ${\bf P}_{A}$ are the mass, position, and momentum, respectively, of the $A$-th nucleus. In Eq.\ (\ref{eq:Hnuc}), the first term is the nuclear kinetic energy (i.e., directly related to the 
nuclear temperature $T$), and the second term is the $T_{el}$-dependent effective potential energy 
\begin{equation}\label{eq:effective-pes} 
U_{\text{eff}}({\bf R}_{1}, ..., {\bf R}_{N}; T_{el}) = - k_{B} T_{el} \ln(\sum_{k} \exp[-\frac{U_{k}({\bf R}_{1}, ..., {\bf R}_{N})}{k_{B} T_{el}}]), 
\end{equation} 
with the $k$-th potential energy $U_{k}({\bf R}_{1}, ..., {\bf R}_{N})$ being given by 
\begin{equation}\label{eq:k-pes} 
U_{k}({\bf R}_{1}, ..., {\bf R}_{N}) = E_{k}({\bf R}_{1}, ..., {\bf R}_{N}) + V_{\text{NN}}({\bf R}_{1}, ..., {\bf R}_{N}), 
\end{equation} 
where the $k$-th eigenvalue $E_{k}({\bf R}_{1}, ..., {\bf R}_{N})$ of the electronic Hamiltonian $\hat{H}_{el}$ (e.g., see Eq.\ (1) of the TAO-AIMD paper \cite{tao-aimd}) and the nuclear-nuclear repulsion energy 
$V_{\text{NN}}({\bf R}_{1}, ..., {\bf R}_{N})$ are both functions of the nuclear positions $\{{\bf R}_{1}, ..., {\bf R}_{N}\}$. The $T_{el}$-dependent effective potential energy function 
$U_{\text{eff}}({\bf R}_{1}, ..., {\bf R}_{N}; T_{el})$ naturally includes the electronic temperature effects. 

Substituting Eq.\ (\ref{eq:k-pes}) into Eq.\ (\ref{eq:effective-pes}) yields 
\begin{equation}\label{eq:effective-pes-1} 
U_{\text{eff}}({\bf R}_{1}, ..., {\bf R}_{N}; T_{el}) = - k_{B} T_{el} \ln(\sum_{k} \exp[-\frac{E_{k}({\bf R}_{1}, ..., {\bf R}_{N})}{k_{B} T_{el}}]) + V_{\text{NN}}({\bf R}_{1}, ..., {\bf R}_{N}). 
\end{equation} 
Based on the canonical ensemble theory \cite{McQuarrie}, the first term in Eq.\ (\ref{eq:effective-pes-1}) is the electronic Helmholtz free energy $F({\bf R}_{1}, ..., {\bf R}_{N}; T_{el})$ at temperature $T_{el}$, 
for fixed nuclear positions $\{{\bf R}_{1}, ..., {\bf R}_{N}\}$. Therefore, Eq.\ (\ref{eq:effective-pes-1}) can also be expressed as 
\begin{equation}\label{eq:effective-pes-1a} 
U_{\text{eff}}({\bf R}_{1}, ..., {\bf R}_{N}; T_{el}) = F({\bf R}_{1}, ..., {\bf R}_{N}; T_{el}) + V_{\text{NN}}({\bf R}_{1}, ..., {\bf R}_{N}). 
\end{equation} 
Accordingly, the electrons are always in thermal equilibrium at temperature $T_{el}$ for fixed nuclear positions $\{{\bf R}_{1}, ..., {\bf R}_{N}\}$. This implies that in this framework, the electrons can respond instantaneously 
to the nuclear motion, similar to the Born-Oppenheimer (BO) approximation \cite{BO-approx}. At zero electronic temperature ($T_{el} = 0\ \text{K}$), $U_{\text{eff}}({\bf R}_{1}, ..., {\bf R}_{N}; T_{el})$ reduces to 
the potential energy function of the electronic GS (also called the GS potential energy function), $U_{0}({\bf R}_{1}, ..., {\bf R}_{N})$: 
\begin{equation}\label{eq:effective-pes-1a1} 
\begin{aligned} 
U_{\text{eff}}({\bf R}_{1}, ..., {\bf R}_{N}; T_{el} = 0\ \text{K}) 
&= F({\bf R}_{1}, ..., {\bf R}_{N}; T_{el} = 0\ \text{K}) + V_{\text{NN}}({\bf R}_{1}, ..., {\bf R}_{N})  \\ 
&= E_{0}({\bf R}_{1}, ..., {\bf R}_{N}) + V_{\text{NN}}({\bf R}_{1}, ..., {\bf R}_{N})  \\ 
&= U_{0}({\bf R}_{1}, ..., {\bf R}_{N}), 
\end{aligned} 
\end{equation} 
where $E_{0}({\bf R}_{1}, ..., {\bf R}_{N})$ (i.e., the lowest eigenvalue of the electronic Hamiltonian $\hat{H}_{el}$) is the electronic GS energy for fixed nuclear positions $\{{\bf R}_{1}, ..., {\bf R}_{N}\}$. 

As a consequence, for fixed nuclear positions $\{{\bf R}_{1}, ..., {\bf R}_{N}\}$, the $T_{el}$-dependent effective potential energy (see Eq.\ (\ref{eq:effective-pes-1a})), obtained with FT-TAO-DFT, is given by 
\begin{equation}\label{eq:TAO-PE} 
U_{\text{eff}}^{\text{FT-TAO-DFT}}({\bf R}_{1}, ..., {\bf R}_{N}; T_{el}) = F_{\text{FT-TAO-DFT}}^{\theta_{el} = k_{B} T_{el}}({\bf R}_{1}, ..., {\bf R}_{N}) + V_{\text{NN}}({\bf R}_{1}, ..., {\bf R}_{N}), 
\end{equation} 
where $F_{\text{FT-TAO-DFT}}^{\theta_{el} = k_{B} T_{el}}({\bf R}_{1}, ..., {\bf R}_{N})$, given by Eq.\ (\ref{eq:tao-energy}), is the electronic Helmholtz free energy of the physical system at the electronic 
temperature $T_{el}$, obtained with FT-TAO-DFT, for the nuclear positions $\{{\bf R}_{1}, ..., {\bf R}_{N}\}$, and $V_{\text{NN}}({\bf R}_{1}, ..., {\bf R}_{N})$ is the corresponding nuclear-nuclear repulsion energy. 
In particular, at zero electronic temperature ($T_{el} = 0\ \text{K}$), 
\begin{equation}\label{eq:TAO-PE2} 
\begin{aligned} 
U_{\text{eff}}^{\text{FT-TAO-DFT}}({\bf R}_{1}, ..., {\bf R}_{N}; T_{el} = 0\ \text{K}) 
&= E_{\text{TAO-DFT}}({\bf R}_{1}, ..., {\bf R}_{N}) + V_{\text{NN}}({\bf R}_{1}, ..., {\bf R}_{N})  \\ 
&= U_{0}^{\text{TAO-DFT}}({\bf R}_{1}, ..., {\bf R}_{N}), 
\end{aligned} 
\end{equation} 
where $E_{\text{TAO-DFT}}({\bf R}_{1}, ..., {\bf R}_{N})$ (e.g., see Eq.\ (35) of the TAO-DFT paper \cite{tao1}) is the electronic GS energy, obtained with TAO-DFT, for the nuclear positions 
$\{{\bf R}_{1}, ..., {\bf R}_{N}\}$, and $U_{0}^{\text{TAO-DFT}}({\bf R}_{1}, ..., {\bf R}_{N})$ is the corresponding GS potential energy. 

As aforementioned, the electrons can respond instantaneously to the nuclear motion in this framework. Accordingly, FT-TAO-DFT can be seamlessly combined with AIMD \cite{Jensen,AIMD2002}, leading to 
the FT-TAO-AIMD method. In FT-TAO-AIMD, the nuclei obey the classical nuclear Hamiltonian (see Eq.\ (\ref{eq:Hnuc})): 
\begin{equation}\label{eq:TAO-H} 
\begin{aligned} 
&\ H_{nuc}({\bf R}_{1}(t), ..., {\bf R}_{N}(t), {\bf P}_{1}(t), ..., {\bf P}_{N}(t); T_{el})  \\ 
=&\ \sum_{A = 1}^{N} \frac{|{\bf P}_{A}(t)|^{2}}{2 M_{A}} + U_{\text{eff}}^{\text{FT-TAO-DFT}}({\bf R}_{1}(t), ..., {\bf R}_{N}(t); T_{el}), 
\end{aligned} 
\end{equation} 
where ${\bf R}_{A}(t)$ and ${\bf P}_{A}(t)$ are the position and momentum, respectively, of the $A$-th nucleus at time $t$. In Eq.\ (\ref{eq:TAO-H}), the first term is the nuclear kinetic energy at time $t$, and the 
second term is the $T_{el}$-dependent effective potential energy (see Eq.\ (\ref{eq:TAO-PE})), obtained with FT-TAO-DFT, for the nuclear positions at time $t$, $\{{\bf R}_{1}(t), ..., {\bf R}_{N}(t)\}$: 
\begin{equation}\label{eq:TAO-PES} 
\begin{aligned} 
&\ U_{\text{eff}}^{\text{FT-TAO-DFT}}({\bf R}_{1}(t), ..., {\bf R}_{N}(t); T_{el})  \\ 
=&\ F_{\text{FT-TAO-DFT}}^{\theta_{el} = k_{B} T_{el}}({\bf R}_{1}(t), ..., {\bf R}_{N}(t)) + V_{\text{NN}}({\bf R}_{1}(t), ..., {\bf R}_{N}(t)), 
\end{aligned} 
\end{equation} 
where $F_{\text{FT-TAO-DFT}}^{\theta_{el} = k_{B} T_{el}}({\bf R}_{1}(t), ..., {\bf R}_{N}(t))$, given by Eq.\ (\ref{eq:tao-energy}), is the electronic Helmholtz free energy of the physical system at the electronic 
temperature $T_{el}$, obtained with FT-TAO-DFT, for the instantaneous nuclear positions $\{{\bf R}_{1}(t), ..., {\bf R}_{N}(t)\}$, and $V_{\text{NN}}({\bf R}_{1}(t), ..., {\bf R}_{N}(t))$ is the corresponding 
nuclear-nuclear repulsion energy. 

According to Eq.\ (\ref{eq:TAO-H}), the nuclei move based on Hamilton's equations of motion on the $T_{el}$-dependent effective potential energy surface generated by FT-TAO-DFT: 
\begin{equation}\label{eq:NE1} 
\dot{{\bf R}}_{A}(t) = \frac{{\bf P}_{A}(t)}{M_{A}} 
\end{equation} 
\begin{equation}\label{eq:NE2} 
\dot{{\bf P}}_{A}(t) = - {\bf \nabla}_{A} U_{\text{eff}}^{\text{FT-TAO-DFT}}({\bf R}_{1}(t), ..., {\bf R}_{N}(t); T_{el}), 
\end{equation} 
where $\dot{{\bf R}}_{A}(t)$ and $\dot{{\bf P}}_{A}(t)$ are the time derivatives of the position and momentum, respectively, of the $A$-th nucleus at time $t$. 
Given the initial nuclear positions $\{{\bf R}_{1}(0), ..., {\bf R}_{N}(0)\}$ and momenta $\{{\bf P}_{1}(0), ..., {\bf P}_{N}(0)\}$, all the future nuclear positions $\{{\bf R}_{1}(t), ..., {\bf R}_{N}(t)\}$ and 
momenta $\{{\bf P}_{1}(t), ..., {\bf P}_{N}(t)\}$ are determined by \Cref{eq:TAO-PES,eq:NE1,eq:NE2}, generating an FT-TAO-AIMD trajectory. 

It is worth mentioning that by construction, the nuclear temperature $T$ (i.e., directly related to the nuclear kinetic energy) is the same as the electronic temperature $T_{el}$ in the FT-TAO-AIMD method. 
By contrast, the nuclear temperature $T$ is generally different from the electronic temperature $T_{el} = 0\ \text{K}$ in the TAO-AIMD method \cite{tao-aimd}. 
However, at very low electronic temperatures ($T_{el} \approx 0\ \text{K}$), 
\begin{equation}\label{eq:effective-pes-tao-aimd} 
\begin{aligned} 
&\ U_{\text{eff}}^{\text{FT-TAO-DFT}}({\bf R}_{1}(t), ..., {\bf R}_{N}(t); T_{el} \approx 0\ \text{K})  \\ 
\approx&\ U_{\text{eff}}^{\text{FT-TAO-DFT}}({\bf R}_{1}(t), ..., {\bf R}_{N}(t); T_{el} = 0\ \text{K})  \\ 
=&\ E_{\text{TAO-DFT}}({\bf R}_{1}(t), ..., {\bf R}_{N}(t)) + V_{\text{NN}}({\bf R}_{1}(t), ..., {\bf R}_{N}(t))  \\ 
=&\ U_{0}^{\text{TAO-DFT}}({\bf R}_{1}(t), ..., {\bf R}_{N}(t)), 
\end{aligned} 
\end{equation} 
where $E_{\text{TAO-DFT}}({\bf R}_{1}(t), ..., {\bf R}_{N}(t))$ (e.g., see Eq.\ (35) of the TAO-DFT paper \cite{tao1}) is the electronic GS energy, obtained with TAO-DFT, for the instantaneous nuclear positions 
$\{{\bf R}_{1}(t), ..., {\bf R}_{N}(t)\}$, and $U_{0}^{\text{TAO-DFT}}({\bf R}_{1}(t), ..., {\bf R}_{N}(t))$ is the corresponding GS potential energy. Accordingly, the classical nuclear Hamiltonian (see Eq.\ (\ref{eq:TAO-H})) 
in FT-TAO-AIMD can be approximately given by the classical nuclear Hamiltonian (e.g., see Eq.\ (4) of the TAO-AIMD paper \cite{tao-aimd}) in TAO-AIMD. Therefore, at very low electronic temperatures 
($T_{el} \approx 0\ \text{K}$), FT-TAO-AIMD can be approximated by TAO-AIMD \cite{tao-aimd}.

\section{FT-TAO-QM/MM}\label{tao-qmmm} 

Here, the FT-TAO-QM/MM method, which combines FT-TAO-DFT and QM/MM \cite{QMMM-1,QMMM-2,QMMM-3}, is developed to offer a cost-effective description of the TE properties of a QM subsystem with 
MR character embedded in an MM environment at finite temperatures. 

In FT-TAO-QM/MM, a physical system is first decomposed into the QM region (e.g., the small, reactive part of the system containing $N$ QM atoms with the coordinates $\{{\bf R}_{1}, ..., {\bf R}_{N}\}$) and the 
MM region (e.g., the surrounding environment containing $K$ MM atoms with the coordinates $\{{\bf X}_{1}, ..., {\bf X}_{K}\}$), wherein the QM region is treated quantum-mechanically by FT-TAO-DFT (i.e., the QM 
potential energy function is computed using FT-TAO-DFT for accuracy), and the MM region is treated classically by molecular mechanics (i.e., a set of empirical potential energy functions (also called a force field) 
is directly adopted for efficiency). 

Without loss of generality, in the present work, the additive scheme \cite{qmmm-review} is adopted to obtain the FT-TAO-QM/MM potential energy function 
\begin{equation}\label{eq:V-TAO-QMMM} 
\begin{aligned} 
&\ U_{\text{FT-TAO-QM/MM}}({\bf R}_{1}, ..., {\bf R}_{N}, {\bf X}_{1}, ..., {\bf X}_{K}; T_{el})  \\ 
=&\ U_{\text{eff}}^{\text{FT-TAO-DFT}}({\bf R}_{1}, ..., {\bf R}_{N}; T_{el}) + V_{\text{MM}}({\bf X}_{1}, ..., {\bf X}_{K}; T_{el})  \\ 
&+ V_{\text{QM-MM}}({\bf R}_{1}, ..., {\bf R}_{N}, {\bf X}_{1}, ..., {\bf X}_{K}; T_{el}), 
\end{aligned} 
\end{equation} 
where 
the interactions between the QM atoms are described by the FT-TAO-DFT $T_{el}$-dependent effective potential energy function $U_{\text{eff}}^{\text{FT-TAO-DFT}}$ (see Eq.\ (\ref{eq:TAO-PE})), 
the interactions between the MM atoms are described by the empirical potential energy function $V_{\text{MM}}$, and 
the interactions between the QM atoms and MM atoms are described by the empirical potential energy function $V_{\text{QM-MM}}$. 
Similar to the QM potential energy function $U_{\text{eff}}^{\text{FT-TAO-DFT}}$, the potential energy functions $V_{\text{MM}}$ and $V_{\text{QM-MM}}$ should generally depend on 
the electronic temperature $T_{el}$. 

At zero electronic temperature ($T_{el} = 0\ \text{K}$), the FT-TAO-QM/MM potential energy function $U_{\text{FT-TAO-QM/MM}}$ (see Eq.\ (\ref{eq:V-TAO-QMMM})) reduces to the TAO-QM/MM (i.e., 
TAO-DFT-based QM/MM) potential energy function $U_{\text{TAO-QM/MM}}$: 
\begin{equation}\label{eq:V-TAO-QMMM2} 
\begin{aligned} 
&\ U_{\text{FT-TAO-QM/MM}}({\bf R}_{1}, ..., {\bf R}_{N}, {\bf X}_{1}, ..., {\bf X}_{K}; T_{el} = 0\ \text{K})  \\ 
=&\ U_{0}^{\text{TAO-DFT}}({\bf R}_{1}, ..., {\bf R}_{N}) + V^{'}_{\text{MM}}({\bf X}_{1}, ..., {\bf X}_{K})  \\ 
&+ V^{'}_{\text{QM-MM}}({\bf R}_{1}, ..., {\bf R}_{N}, {\bf X}_{1}, ..., {\bf X}_{K}) \\ 
=&\ U_{\text{TAO-QM/MM}}({\bf R}_{1}, ..., {\bf R}_{N}, {\bf X}_{1}, ..., {\bf X}_{K}), 
\end{aligned} 
\end{equation} 
where $U_{0}^{\text{TAO-DFT}}$ is the TAO-DFT GS potential energy function (see Eq.\ (\ref{eq:TAO-PE2})), $V^{'}_{\text{MM}} \equiv V_{\text{MM}}$ (with $T_{el} = 0\ \text{K}$), and 
$V^{'}_{\text{QM-MM}} \equiv V_{\text{QM-MM}}$ (with $T_{el} = 0\ \text{K}$). 

In short, FT-TAO-QM/MM is well-suited for the study of an MR system embedded in an MM environment at finite temperatures. For example, FT-TAO-QM/MM can be employed to simulate the experimental IR 
spectra of $n$-acene obtained with the matrix-isolation techniques \cite{matrix-isolation-1,matrix-isolation-2,matrix-isolation-book}, wherein $n$-acene (treated by FT-TAO-DFT) and the rare gas atoms (treated 
by MM) are co-deposited at very low temperatures \cite{acene-EXP-IR-1,acene-EXP-IR-2}.

\section{Computational Details}\label{sec:comp} 

The LDA xc$\theta$ free energy functional $F_{\text{xc}\theta}^{\text{LDA},\theta_{el}}[\rho_{\alpha},\rho_{\beta}]$ (i.e., a combined LDA xc \cite{ftlda} and $\theta$-dependent \cite{tao1,perrot-1979} free energy 
functional, see Eq.\ (\ref{eq:LDAxctheta})) with the fictitious temperature $\theta$ = 7 mhartree (i.e., the optimal system-independent $\theta$ for TAO-LDA (i.e., FT-TAO-LDA at $\theta_{el} = 0$)) \cite{tao1} is 
adopted for all the FT-TAO-DFT, FT-TAO-AIMD, and FT-TAO-QM/MM calculations. All calculations are spin-restricted, and are performed with \textsf{Q-Chem 6} \cite{Q-Chem5}, using the 6-31G(d) basis set with 
a numerical grid of 75 Euler-Maclaurin radial grid points and 302 Lebedev angular grid points. 

\begin{figure} 
\includegraphics[scale=0.36]{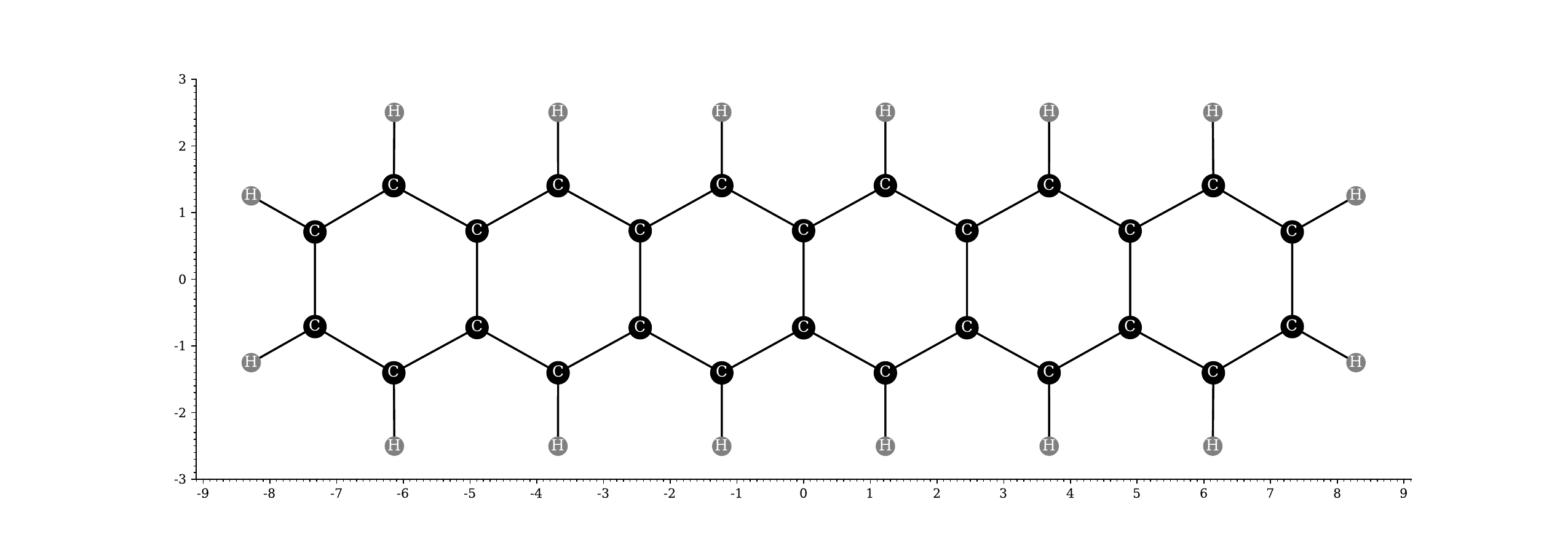} 
\caption{\label{fig:6-acene-geo} 
FT-TAO-DFT optimized geometry (in {\AA}) of 6-acene in vacuum at the electronic temperature $T_{el}$ = 1000 K.} 
\end{figure}

Each FT-TAO-DFT calculation is performed at the geometry of $n$-acene in vacuum, corresponding to the minimum of $U_{\text{eff}}^{\text{FT-TAO-DFT}}$ (see Eq.\ (\ref{eq:TAO-PE})), which is hereafter referred 
to as the FT-TAO-DFT optimized geometry of $n$-acene in vacuum at the electronic temperature $\theta_{el} = k_{B} T_{el}$ (see \Cref{fig:6-acene-geo}), wherein the nuclear kinetic energy is ignored. The 
symmetrized von Neumann entropy, active orbital occupation numbers, and IR spectra of $n$-acene ($n$ = 2--6) in vacuum at the electronic temperature $T_{el}$ = 0 K, 300 K, and 1000 K are computed to examine 
the electronic temperature effects. For $n$-acene (with $N_{el}$ electrons), the $(N_{el}/2)$-th orbital is defined as the highest occupied molecular orbital (HOMO), the $(N_{el}/2+1)$-th orbital is defined as the lowest 
unoccupied molecular orbital (LUMO), and so on \cite{tao1,tao3,tao-gnr}. Besides, the orbitals with an occupation number between 0.2 and 1.8 are regarded as the active orbitals. 

Here, the nuclear kinetic energy (i.e., directly related to the nuclear temperature $T$) is reintroduced, and the nuclear dynamics is described by FT-TAO-AIMD. 
FT-TAO-AIMD simulations are performed to compute the symmetrized von Neumann entropy, active orbital occupation numbers, and IR spectra of $n$-acene ($n$ = 2--6) in vacuum at the nuclear temperature 
$T$ = $T_{el}$ = 1000 K. For each FT-TAO-AIMD simulation, the time step for the integration of the equations of motion is chosen as 20 a.u. ($\approx$ 0.484 fs). The simulation is initialized from the FT-TAO-DFT 
optimized geometry of $n$-acene in vacuum at $T_{el}$ = 1000 K, with the initial nuclear velocities being randomly generated using the Maxwell-Boltzmann (MB) distribution at $T$ = 1000 K. The system then evolves 
in the canonical ensemble with the aid of Nos\'{e}-Hoover (NH) chain thermostat \cite{NH-chain} (using the default settings of \textsf{Q-Chem}) for about 10.2 ps. To prevent its interference with the 
dynamics \cite{thermostat-infer}, the NH chain thermostat is subsequently removed, and the FT-TAO-AIMD simulation proceeds in the microcanonical ensemble for 10500 time steps ($\approx$ 5.1 ps) to equilibrate 
the system. After equilibration, the simulation continues in the microcanonical ensemble for 42000 time steps ($\approx$ 20.3 ps) to collect relevant data along the FT-TAO-AIMD equilibrated trajectory (with the 
average nuclear temperature being $1000 \pm 50$ K). To sample different regions in the phase space, since the initial nuclear velocities for each FT-TAO-AIMD simulation are randomly generated using the MB 
distribution at $T$ = 1000 K, the aforementioned processes are repeated to produce four different FT-TAO-AIMD equilibrated trajectories ($\approx$ 20.3 ps per trajectory). Along each FT-TAO-AIMD equilibrated 
trajectory, the IR spectrum is computed using the \textsf{TRAVIS} program package \cite{Brehm-Kirchner-2011,Thomas-2013,Thomas-2015,Brehm-2020}. 

\begin{figure} 
\subfigure 
{\includegraphics[scale=0.6]{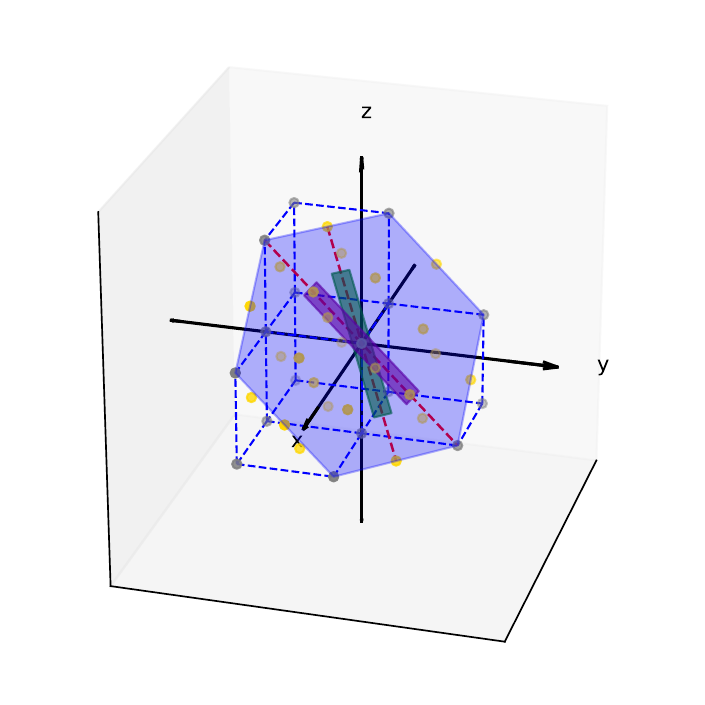}(1)} 
\subfigure 
{\includegraphics[scale=0.6]{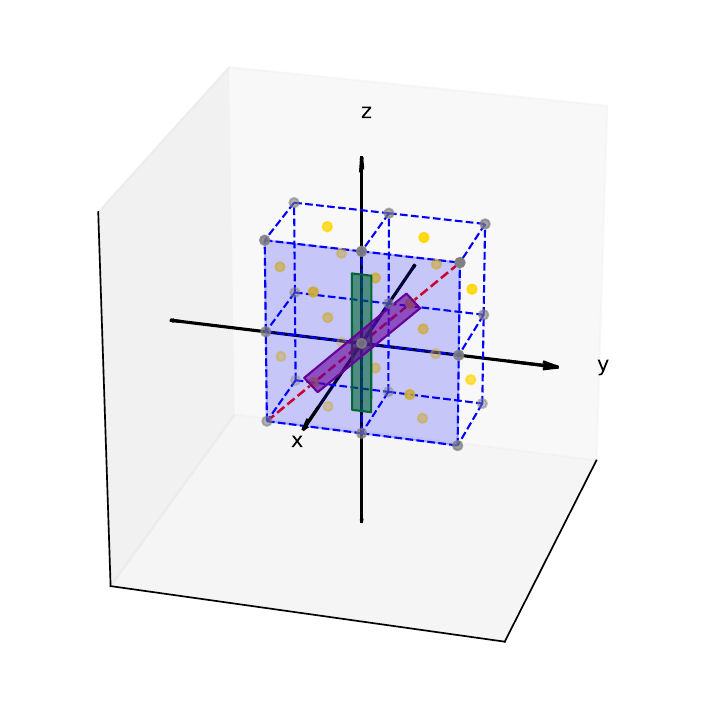}(2)} 
\subfigure 
{\includegraphics[scale=0.6]{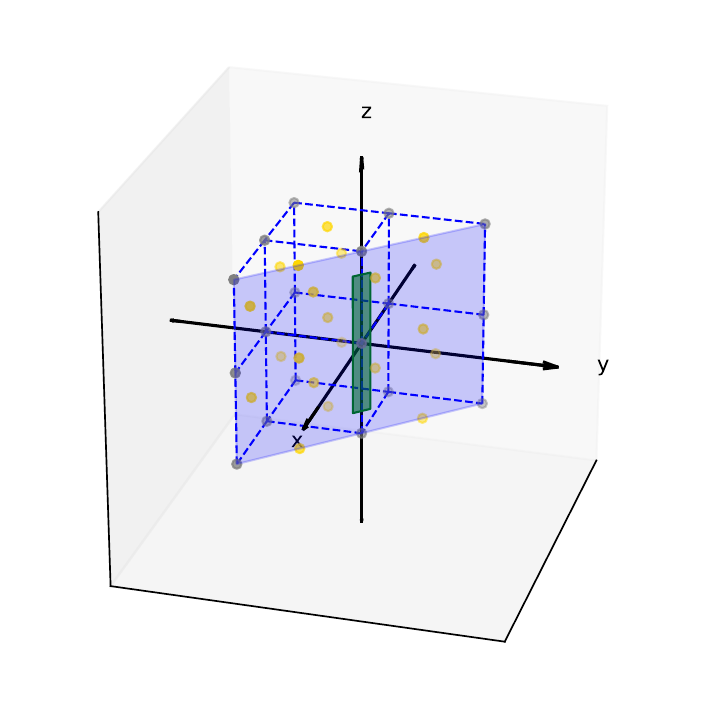}(3)} 
\caption{\label{fig:position} 
Scratch of $n$-acene inserted into an Ar matrix. 
The green or purple rectangle in each subfigure represents the surface where $n$-acene is placed, including 
(1) 1a (green) and 1b (purple) on the (111) plane, 
(2) 2a (green) and 2b (purple) on the (100) plane, and 
(3) 3a (green) on the (110) plane. 
The grey dots and yellow dots represent Ar atoms. 
The inserted $n$-acene and the Ar box are centered at the same geometric point.} 
\end{figure}

To mimic the experimental environment (i.e., $n$-acene in an Ar matrix), we construct an Ar box (containing 2457 Ar atoms) with a lattice parameter $a_{0}$ = 5.320 {\AA}, corresponding to the face-centered cubic 
structure of an Ar matrix at very low temperatures \cite{argon-exp}. As illustrated in \Cref{fig:position}, $n$-acene is inserted into the Ar matrix at five different positions (1a, 1b, 2a, 2b, and 3a). Since the temperature 
considered in the matrix-isolation technique is extremely low ($\approx$ 10 K) \cite{acene-EXP-IR-1,acene-EXP-IR-2}, the electronic temperature $T_{el}$ = 0 K is adopted here. For the system considered (i.e., 
$n$-acene in an Ar matrix), FT-TAO-QM/MM is well-suited, wherein $n$-acene is chosen as the QM region (i.e., treated quantum-mechanically by FT-TAO-DFT for accuracy, due to the potential MR character of 
$n$-acene \cite{tao1,tao2,tao3,dmrg}), and the Ar matrix is chosen as the MM region (i.e., treated classically by molecular mechanics for efficiency). 
Since no covalent bonds exist between the MM atoms (i.e., dominated by van der Waals interactions), the MM potential energy function is described by 
\begin{equation}\label{eq:V-MM} 
V^{'}_{\text{MM}}({\bf X}_{1}, ..., {\bf X}_{K}) =  \sum_{I = 1}^{K} \sum_{J > I}^{K} v_{\text{LJ}}({\bf X}_{I}, \sigma_{I}, \epsilon_{I}; {\bf X}_{J}, \sigma_{J}, \epsilon_{J}), 
\end{equation} 
where 
\begin{equation}\label{eq:LJ-pot} 
v_{\text{LJ}}({\bf X}_{I}, \sigma_{I}, \epsilon_{I}; {\bf X}_{J}, \sigma_{J}, \epsilon_{J}) 
= 4 \epsilon_{IJ} \bigg[\bigg(\frac{\sigma_{IJ}}{|{\bf X}_{I} - {\bf X}_{J}|}\bigg)^{12} - \bigg(\frac{\sigma_{IJ}}{|{\bf X}_{I} - {\bf X}_{J}|}\bigg)^{6}\bigg] 
\end{equation} 
is the Lennard-Jones (LJ) potential \cite{Jensen} for a pair of atom $I$ and atom $J$ (both in the MM region), with $\sigma_{IJ} = (\sigma_{I} \sigma_{J})^{1/2}$ and $\epsilon_{IJ} = (\epsilon_{I} \epsilon_{J})^{1/2}$ 
being computed using the atomic LJ parameters $\sigma_{I}$, $\sigma_{J}$, $\epsilon_{I}$, and $\epsilon_{J}$ \cite{OPLSAA-1}. Here, geometric combining rules for the LJ parameters are adopted. 
Similarly, since no covalent bonds exist between the QM atoms and MM atoms (i.e., dominated by van der Waals interactions), the QM-MM potential energy function is described by 
\begin{equation}\label{eq:H-QM-MM} 
V^{'}_{\text{QM-MM}}({\bf R}_{1}, ..., {\bf R}_{N}, {\bf X}_{1}, ..., {\bf X}_{K}) = \sum_{A = 1}^{N} \sum_{I = 1}^{K} v_{\text{LJ}}({\bf R}_{A}, \sigma_{A}, \epsilon_{A}; {\bf X}_{I}, \sigma_{I}, \epsilon_{I}), 
\end{equation} 
where 
\begin{equation}\label{eq:LJ-pot2} 
v_{\text{LJ}}({\bf R}_{A}, \sigma_{A}, \epsilon_{A}; {\bf X}_{I}, \sigma_{I}, \epsilon_{I}) 
= 4 \epsilon_{AI} \bigg[\bigg(\frac{\sigma_{AI}}{|{\bf R}_{A} - {\bf X}_{I}|}\bigg)^{12} - \bigg(\frac{\sigma_{AI}}{|{\bf R}_{A} - {\bf X}_{I}|}\bigg)^{6}\bigg], 
\end{equation} 
is the LJ potential for a pair of atom $A$ (in the QM region) and atom $I$ (in the MM region), with $\sigma_{AI} = (\sigma_{A} \sigma_{I})^{1/2}$ and $\epsilon_{AI} = (\epsilon_{A} \epsilon_{I})^{1/2}$ being computed 
using the atomic LJ parameters $\sigma_{A}$, $\sigma_{I}$, $\epsilon_{A}$, and $\epsilon_{I}$ \cite{OPLSAA-1}. All the atomic LJ parameters (i.e., $\epsilon$ and $r_{\text{min}} \equiv 2^{1/6} \sigma$) are taken from 
the all-atom optimized potentials for liquid simulations (OPLS-AA) force field \cite{OPLSAA-1,OPLSAA-2}, which are listed in \Cref{table:LJ-coeff}. Besides, for computational efficiency, we follow the strategy of Sander 
{\it et al.} in a recent QM/MM study \cite{insert} that the Ar atoms lying too close to $n$-acene (e.g., within the half of the sum of $r_{\text{min}}$ values of the nearby atoms) are manually removed. 
Each FT-TAO-QM/MM calculation is performed at the geometry of $n$-acene in the Ar matrix, corresponding to the minimum of $U_{\text{FT-TAO-QM/MM}}$ (see Eq.\ (\ref{eq:V-TAO-QMMM})), which is hereafter 
referred to as the FT-TAO-QM/MM optimized geometry of $n$-acene in the Ar matrix at $T_{el}$ = 0 K. The symmetrized von Neumann entropy, active orbital occupation numbers, and IR spectra of $n$-acene ($n$ = 2--6) 
inserted into the Ar matrix at various positions (1a, 1b, 2a, 2b, and 3a) at $T_{el}$ = 0 K are computed to examine the effects of the Ar matrix, when compared with the corresponding results in vacuum at $T_{el}$ = 0 K. 

Normal mode analysis (NMA) \cite{In61,In62,NMA,tao-aimd} is a commonly used method to calculate the vibrational frequencies and intensities of molecules in the harmonic approximation at absolute zero. Here, NMA is 
employed to compute the IR spectrum of $n$-acene at the electronic temperature $T_{el}$. The NMA-based IR spectrum is obtained with FT-TAO-DFT for $n$-acene in vacuum, and is obtained with FT-TAO-QM/MM for 
$n$-acene in an Ar matrix. To perform the NMA, the computation of nuclear second derivatives of energy (i.e., the nuclear Hessian) at the optimized molecular geometry is needed. Because analytical nuclear Hessians 
for both FT-TAO-DFT and FT-TAO-QM/MM remain unavailable in \textsf{Q-Chem}, numerical nuclear Hessians are computed using finite differences of analytical nuclear gradients (with a step size of 0.001 {\AA}) for all 
the NMA calculations. For all the FT-TAO-QM/MM calculations, given the large number of MM atoms, it is prohibitively expensive to diagonalize the full mass-weighted Hessian matrix. To address this issue, the partial 
Hessian vibrational analysis (PHVA) approach \cite{PHVA,PHVA-1} is adopted, namely only the QM-QM block of the mass-weighted Hessian matrix is diagonalized.

\section{Results and Discussion}\label{sec:results} 

\subsection{Acenes in vacuum}\label{sec:n-acene} 

\subsubsection{Radical nature} 

We first examine the electronic temperature effects on the symmetrized von Neumann entropy \cite{tao2,tao3,tao-aimd} 
\begin{equation}\label{eq:RSvn} 
S_{\text{vN}} = - \sum_{i=1}^{\infty} \bigg\lbrace \frac{f_{i}}{2}\ \text{ln}\bigg( \frac{f_{i}}{2} \bigg) + \bigg( 1-\frac{f_{i}}{2} \bigg)\ \text{ln}\bigg( 1-\frac{f_{i}}{2} \bigg) \bigg\rbrace, 
\end{equation} 
and active orbital occupation numbers of $n$-acene ($n$ = 2--6) in vacuum, by performing FT-TAO-DFT calculations at the electronic temperature $T_{el}$ = 0 K, 300 K, and 1000 K. Here, $f_{i}$ (i.e., a value 
between 0 and 2) is the occupation number of the $i$-th orbital, obtained with FT-TAO-DFT. Here, each calculation is performed at the FT-TAO-DFT optimized geometry of $n$-acene in vacuum at $T_{el}$ (i.e., 
the nuclear kinetic energy is ignored). For brevity, $n$-acene in vacuum may hereafter be called $n$-acene. 

\begin{figure} 
\includegraphics[scale=0.36]{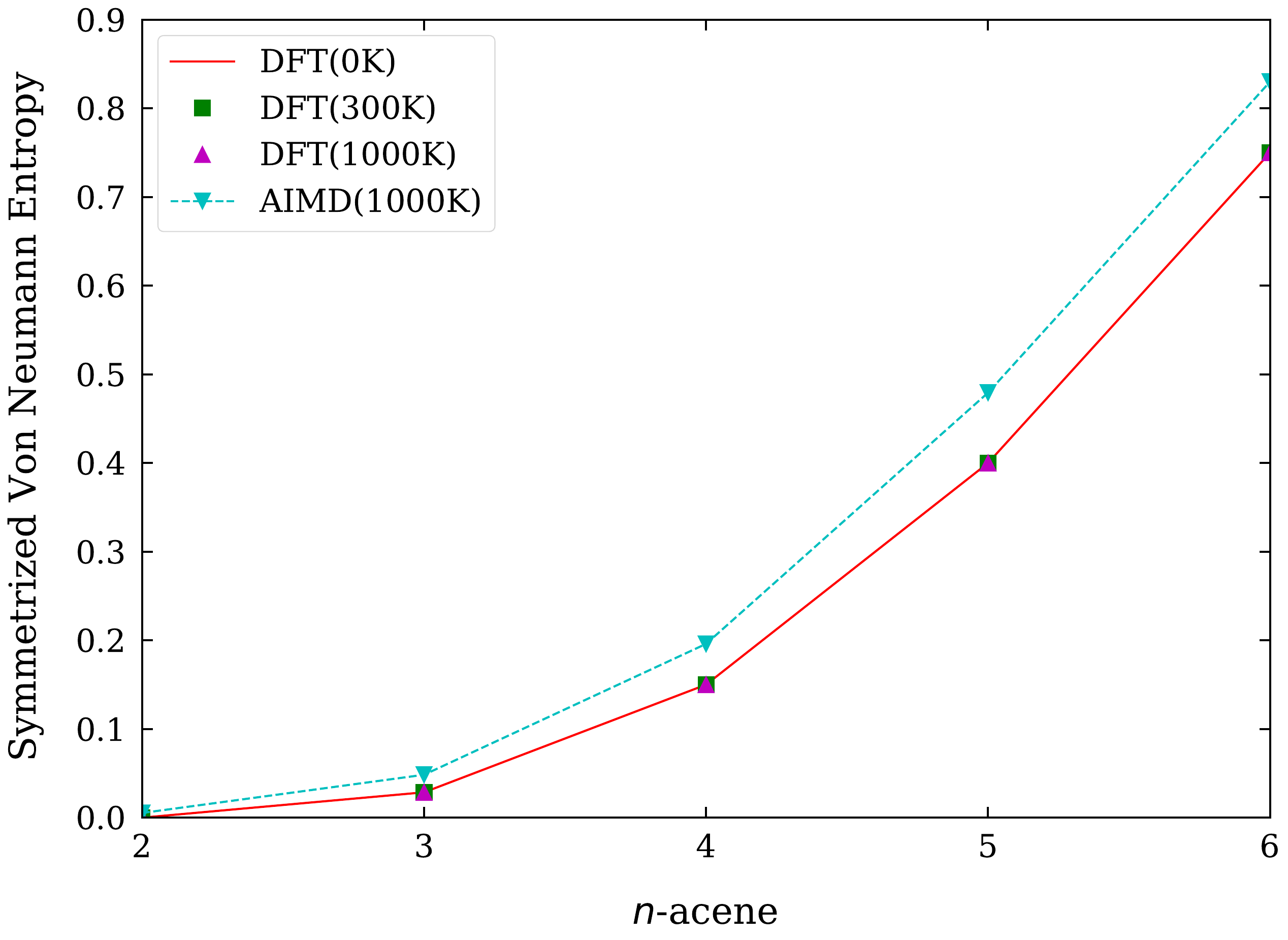} 
\caption{\label{fig:acene-SVN} 
Symmetrized von Neumann entropy ($S_{\text{vN}}$) of $n$-acene in vacuum at the electronic temperature $T_{el}$ = 0 K, 300 K, and 1000 K, computed using FT-TAO-DFT. 
For comparison, the corresponding FT-TAO-AIMD average values ($\overline{S_{\text{vN}}}$ at 1000 K) are also shown.} 
\end{figure}

\begin{figure} 
\includegraphics[scale=0.32]{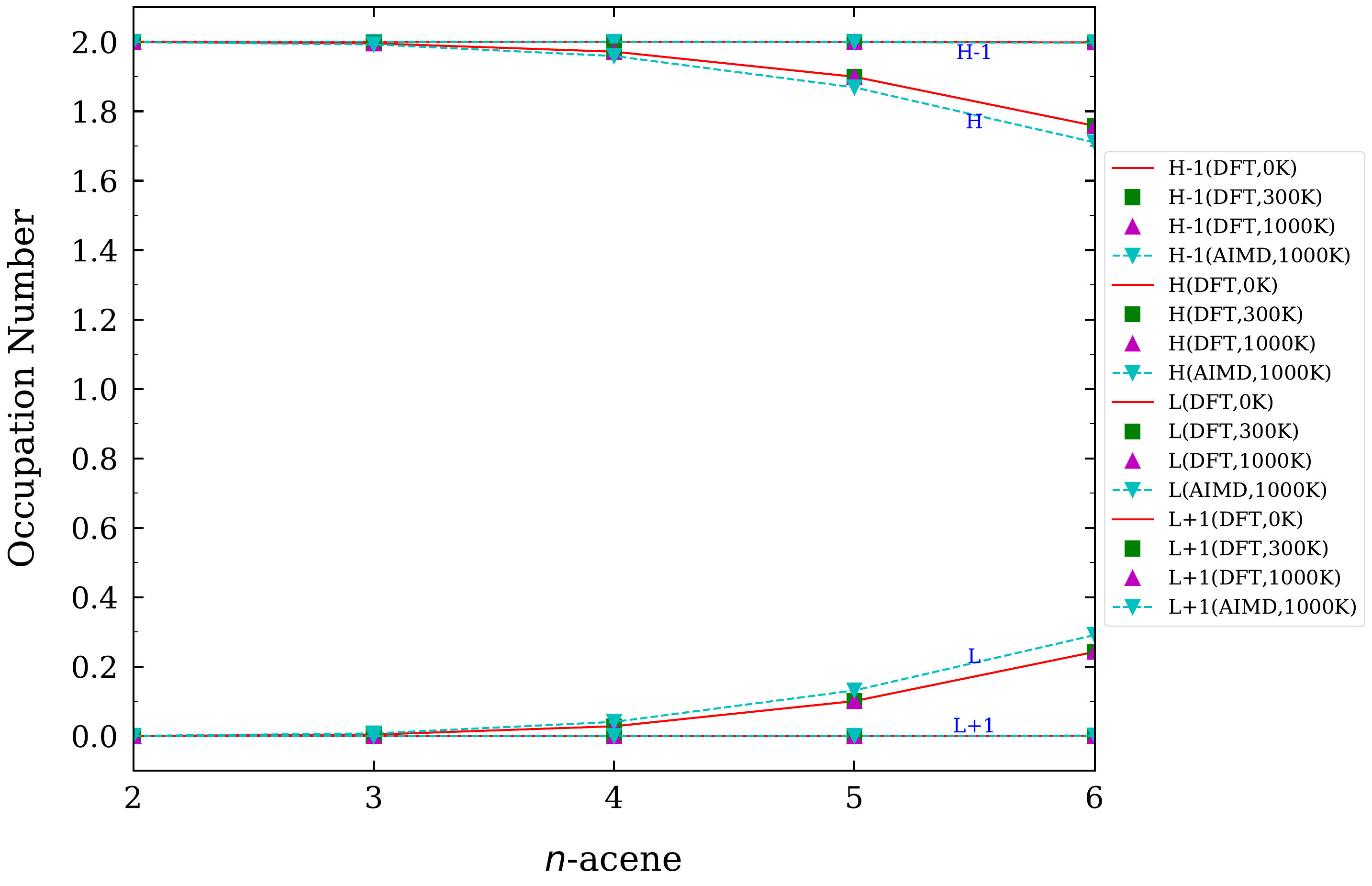} 
\caption{\label{fig:acene-OCC} 
Active orbital occupation numbers ($f_{\text{H-1}}$, $f_{\text{H}}$, $f_{\text{L}}$, and $f_{\text{L+1}}$) of $n$-acene in vacuum at the electronic temperature $T_{el}$ = 0 K, 300 K, and 1000 K, 
computed using FT-TAO-DFT. The HOMO/LUMO is denoted as the H/L for brevity. For comparison, the corresponding 
FT-TAO-AIMD average values ($\overline{f_{\text{H-1}}}$, $\overline{f_{\text{H}}}$, $\overline{f_{\text{L}}}$, and $\overline{f_{\text{L+1}}}$ at 1000 K) are also shown.} 
\end{figure}

The symmetrized von Neumann entropy $S_{\text{vN}}$ (see \Cref{fig:acene-SVN}) and active orbital occupation numbers (see \Cref{fig:acene-OCC}) of $n$-acene are rather insensitive to electronic temperature 
changes, indicating that the electronic temperature effects on these properties are very minor. This also implies that the electronic thermal ensembles of $n$-acene at $T_{el} \le 1000$ K are mainly contributed by 
the electronic GS (i.e., the electronic thermal ensemble at $T_{el}$ = 0 K). Accordingly, similar to the GS counterparts at $T_{el}$ = 0 K, for $n$-acene at $T_{el} \le 1000$ K, the radical nature can be assessed by 
the $S_{\text{vN}}$ \cite{tao2,tao3}, and the orbital occupation numbers can be approximately regarded as the GS NOONs \cite{tao1,Lowdin-1955,Lowdin-1956}. 

For the smaller $n$-acenes ($n$ = 2--5) at $T_{el} \le 1000$ K, all the orbital occupation numbers are very close to either 0 (fully empty) or 2 (fully occupied), leading to very small $S_{\text{vN}}$ values. This 
suggests that the smaller $n$-acenes ($n$ = 2--5) at $T_{el} \le 1000$ K should possess non-radical nature. By contrast, for 6-acene at $T_{el} \le 1000$ K, the HOMO and LUMO occupation numbers noticeably 
deviate from 0 and 2 (e.g., between 0.2 and 1.8), leading to the larger $S_{\text{vN}}$. Therefore, 6-acene at $T_{el} \le 1000$ K should possess noticeable di-radical nature. 

The nuclear temperature $T$ is directly related to the nuclear kinetic energy in FT-TAO-AIMD. Therefore, we further investigate the nuclear temperature effects on the symmetrized von Neumann entropy and active 
orbital occupation numbers of $n$-acene, by performing FT-TAO-AIMD simulations at $T$ = 1000 K (with $T_{el}$ = $T$). Since the electronic thermal ensembles of $n$-acene at $T_{el} \le 1000$ K are mainly 
contributed by the electronic GS at $T_{el}$ = 0 K, FT-TAO-AIMD at $T \le 1000$ K (with $T_{el}$ = $T$) can be approximately given by TAO-AIMD at the same $T$ (with $T_{el}$ = 0 K) \cite{tao-aimd} (see 
Eq.\ (\ref{eq:effective-pes-tao-aimd})). Accordingly, similar to our previous TAO-AIMD study \cite{tao-aimd}, the instantaneous radical nature of $n$-acene at $T$ = 1000 K along each FT-TAO-AIMD equilibrated 
trajectory is examined by the instantaneous symmetrized von Neumann entropy 
\begin{equation}\label{eq:Svn} 
S_{\text{vN}}(t) = - \sum_{i=1}^{\infty} \bigg\lbrace \frac{f_{i}(t)}{2}\ \text{ln}\bigg( \frac{f_{i}(t)}{2} \bigg) + \bigg( 1-\frac{f_{i}(t)}{2} \bigg)\ \text{ln}\bigg( 1-\frac{f_{i}(t)}{2} \bigg) \bigg\rbrace, 
\end{equation} 
where $f_{i}(t)$ is the occupation number of the $i$-th orbital, obtained with FT-TAO-DFT, at the instantaneous nuclear positions $\{{\bf R}_{1}(t), ..., {\bf R}_{N}(t)\}$. Besides, along the FT-TAO-AIMD equilibrated 
trajectory, the time average of $S_{\text{vN}}(t)$ is computed using 
\begin{equation}\label{eq:TASvn} 
\overline{S_{\text{vN}}} = \frac{1}{\tau} \int_{0}^{\tau} S_{\text{vN}}(t) dt, 
\end{equation} 
and the time average of $f_{i}(t)$ is computed using 
\begin{equation}\label{eq:TAfi} 
\overline{f_{i}} = \frac{1}{\tau} \int_{0}^{\tau} f_{i}(t) dt, 
\end{equation} 
with $\tau$ being the total time duration of the FT-TAO-AIMD equilibrated trajectory. The reported $\overline{S_{\text{vN}}}$ or $\overline{f_{i}}$ of $n$-acene is an average over four different FT-TAO-AIMD equilibrated 
trajectories. 

As $n$ increases, the $\overline{S_{\text{vN}}}$ (see \Cref{fig:acene-SVN}) and $\overline{f_{i}}$ (see \Cref{fig:acene-OCC}) values of $n$-acene at $T$ = $T_{el}$ = 1000 K, obtained with FT-TAO-AIMD simulations, 
increasingly deviate from the $S_{\text{vN}}$ and $f_{i}$ values, respectively, of $n$-acene at $T_{el}$ = 1000 K, obtained with FT-TAO-DFT (i.e., the corresponding values where the nuclear kinetic energy is ignored). 
This suggests that on average, the radical nature of $n$-acene obtained with FT-TAO-AIMD simulations is enhanced mainly due to the nuclear motion (e.g., molecular vibrations) at $T$ = 1000 K, and the enhancement 
of radical nature increases with $n$. 

As aforementioned, for $n$-acene, FT-TAO-AIMD simulations at $T$ = 1000 K can be approximately regarded as TAO-AIMD simulations at $T$ = 1000 K, which on average, indeed yield the more enhanced radical nature 
of $n$-acene, when compared with the corresponding radical nature obtained with TAO-AIMD simulations at $T$ = 300 K (e.g., see Figures 2 and 4 in the TAO-AIMD paper \cite{tao-aimd}). With increasing nuclear 
temperature $T$, increased nuclear motion (e.g., molecular vibrations) can cause chemical bonds to stretch, distort, and vibrate more strongly, yielding on average, the more enhanced radical nature of $n$-acene. 

\begin{figure} 
\includegraphics[scale=0.32]{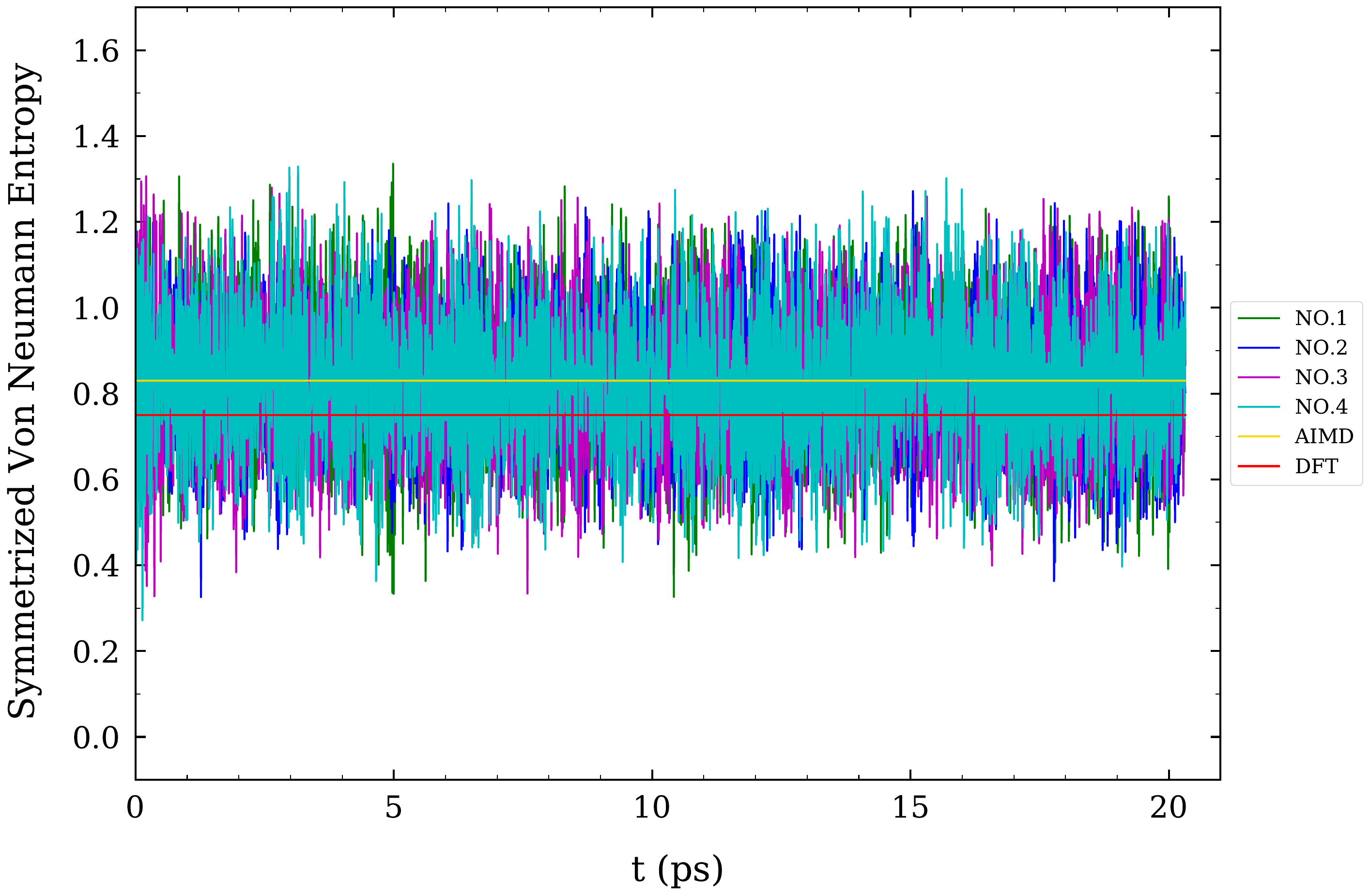} 
\caption{\label{fig:6-acene-NVE-1000K-SVN} 
Instantaneous symmetrized von Neumann entropy ($S_{\text{vN}}(t)$) of 6-acene in vacuum, obtained from four different FT-TAO-AIMD equilibrated trajectories (No.1 to No.4) at 1000 K. For comparison, the 
FT-TAO-AIMD average value ($\overline{S_{\text{vN}}}$ at 1000 K) and 
FT-TAO-DFT value ($S_{\text{vN}}$ at the electronic temperature $T_{el}$ = 1000 K) are also shown.} 
\end{figure}

\begin{figure} 
\includegraphics[scale=0.32]{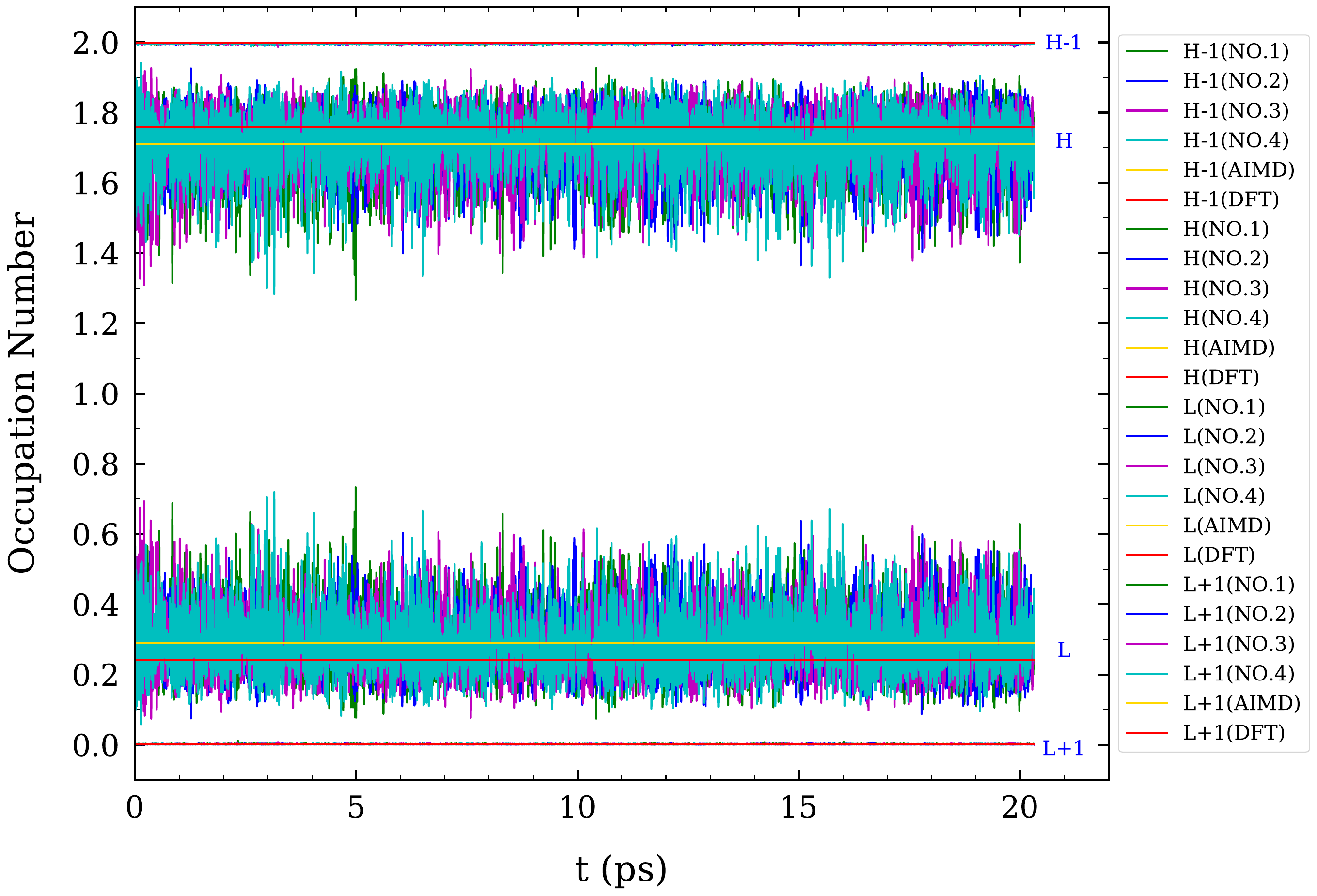} 
\caption{\label{fig:6-acene-NVE-1000K-OCC} 
Instantaneous active orbital occupation numbers ($f_{\text{H-1}}(t)$, $f_{\text{H}}(t)$, $f_{\text{L}}(t)$, and $f_{\text{L+1}}(t)$) of 6-acene in vacuum, obtained from four different FT-TAO-AIMD 
equilibrated trajectories (No.1 to No.4) at 1000 K. The HOMO/LUMO is denoted as the H/L for brevity. For comparison, the 
FT-TAO-AIMD average values ($\overline{f_{\text{H-1}}}$, $\overline{f_{\text{H}}}$, $\overline{f_{\text{L}}}$, and $\overline{f_{\text{L+1}}}$ at 1000 K) and 
FT-TAO-DFT values ($f_{\text{H-1}}$, $f_{\text{H}}$, $f_{\text{L}}$, and $f_{\text{L+1}}$ at the electronic temperature $T_{el}$ = 1000 K) are also shown.} 
\end{figure}

Along each FT-TAO-AIMD equilibrated trajectory, the instantaneous $S_{\text{vN}}(t)$ (see \Cref{fig:6-acene-NVE-1000K-SVN}) and $f_{i}(t)$ (see \Cref{fig:6-acene-NVE-1000K-OCC}) values of 6-acene greatly fluctuate 
around the $\overline{S_{\text{vN}}}$ and $\overline{f_{i}}$ values, respectively (i.e., the corresponding FT-TAO-AIMD averages at 1000 K) over time, due to the nuclear motion at $T$ = 1000 K (also see the Supporting 
Information (SI) for the corresponding results of $n$-acene ($n$ = 2--5)). Therefore, it is essential to perform computationally efficient AIMD simulations that can adequately describe the instantaneous radical nature of 
6-acene at 1000 K, well justifying the employment of FT-TAO-AIMD in the present study.

\subsubsection{IR spectra} 

The IR spectra of polycyclic aromatic hydrocarbons (PAHs) have recently received considerable attention, since PAHs can be potential candidates for the unidentified IR (UIR) emission bands from interstellar space 
\cite{PAH-IR-E,PAH-IR-T,PAH-IR-R}. In particular, $n$-acenes represent a class of PAHs with possible radical nature. To adequately predict the IR spectra of $n$-acenes, TAO-DFT-based NMA and TAO-AIMD \cite{tao-aimd} 
have been recently employed. The IR spectra of $n$-acenes obtained with TAO-AIMD simulations at $T$ = 300 K are qualitatively consistent with the available experimental data \cite{tao-aimd}. 

However, free PAHs can easily reach temperatures of around 1000 K by absorbing an ultraviolet photon, and subsequently relaxing through IR emission in interstellar space \cite{PAH-IR-T,PAH-IR-R}. To properly simulate 
such conditions, we compute the IR spectra of $n$-acenes at 1000 K using both FT-TAO-DFT-based NMA and FT-TAO-AIMD. 

\begin{figure} 
\includegraphics[scale=0.32]{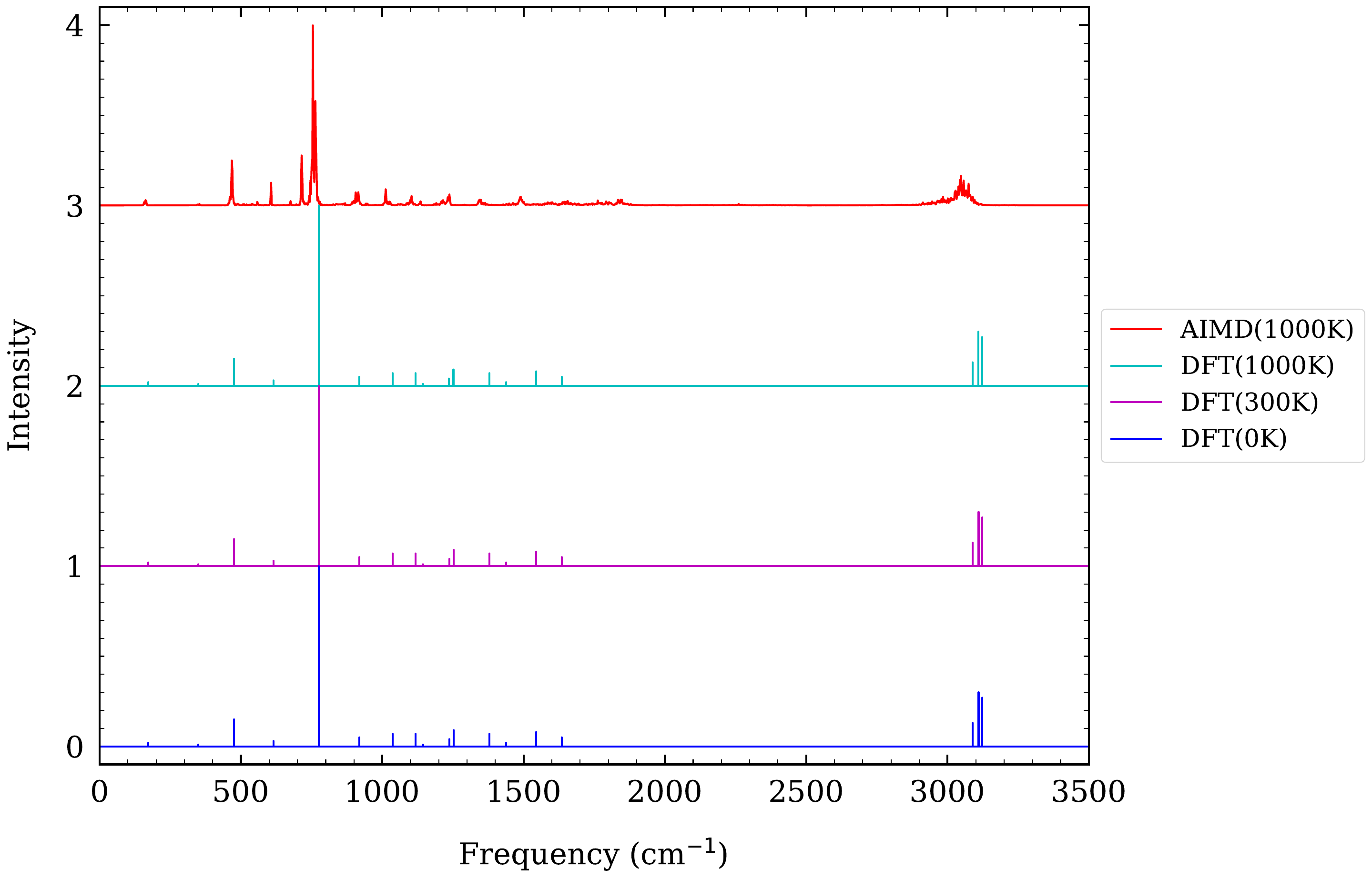} 
\caption{\label{fig:2-acene-NVE-1000K_MD} 
IR spectra of 2-acene in vacuum at the electronic temperature $T_{el}$ = 0 K, 300 K, and 1000 K, computed using FT-TAO-DFT (via NMA). 
For comparison, the IR spectrum obtained with FT-TAO-AIMD simulations at 1000 K is also shown. 
The IR spectra are normalized to have a maximum intensity of one, and are vertically offset from each other by the same value for clarity.} 
\end{figure}

\begin{figure} 
\includegraphics[scale=0.32]{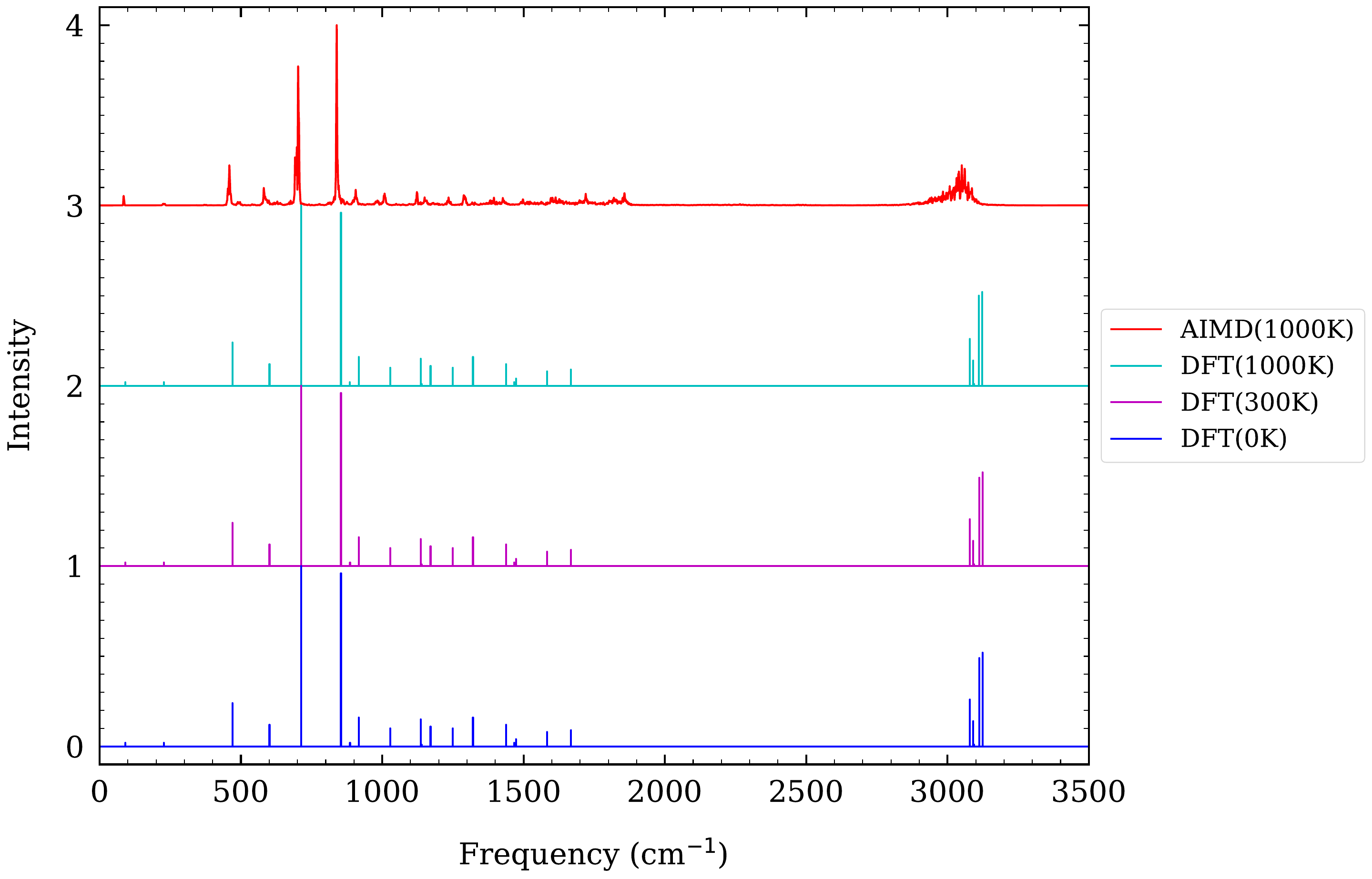} 
\caption{\label{fig:3-acene-NVE-1000K_MD} 
IR spectra of 3-acene in vacuum at the electronic temperature $T_{el}$ = 0 K, 300 K, and 1000 K, computed using FT-TAO-DFT (via NMA). 
For comparison, the IR spectrum obtained with FT-TAO-AIMD simulations at 1000 K is also shown. 
The IR spectra are normalized to have a maximum intensity of one, and are vertically offset from each other by the same value for clarity.} 
\end{figure}

\begin{figure} 
\includegraphics[scale=0.32]{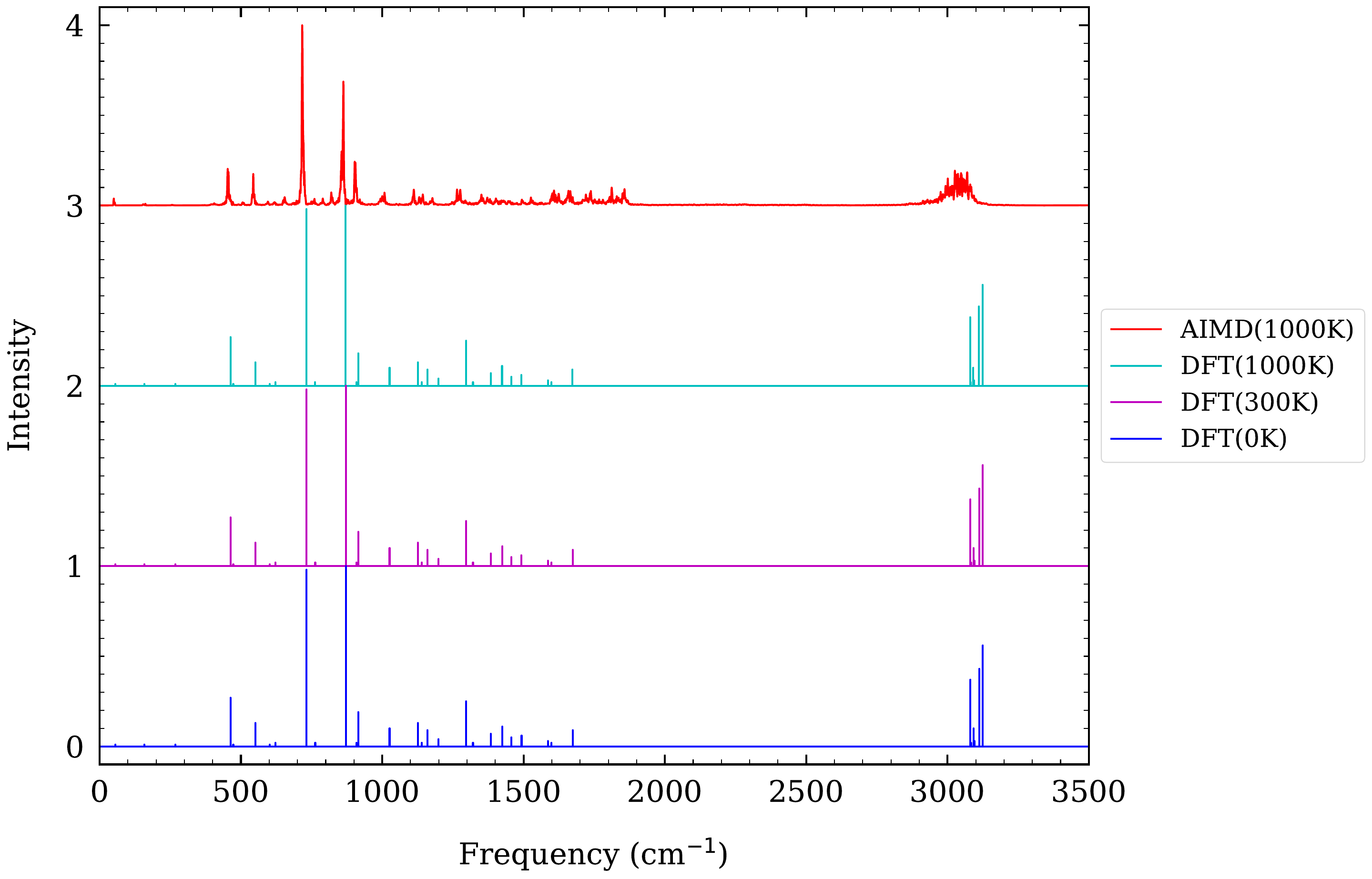} 
\caption{\label{fig:4-acene-NVE-1000K_MD} 
IR spectra of 4-acene in vacuum at the electronic temperature $T_{el}$ = 0 K, 300 K, and 1000 K, computed using FT-TAO-DFT (via NMA). 
For comparison, the IR spectrum obtained with FT-TAO-AIMD simulations at 1000 K is also shown. 
The IR spectra are normalized to have a maximum intensity of one, and are vertically offset from each other by the same value for clarity.} 
\end{figure}

\begin{figure} 
\includegraphics[scale=0.32]{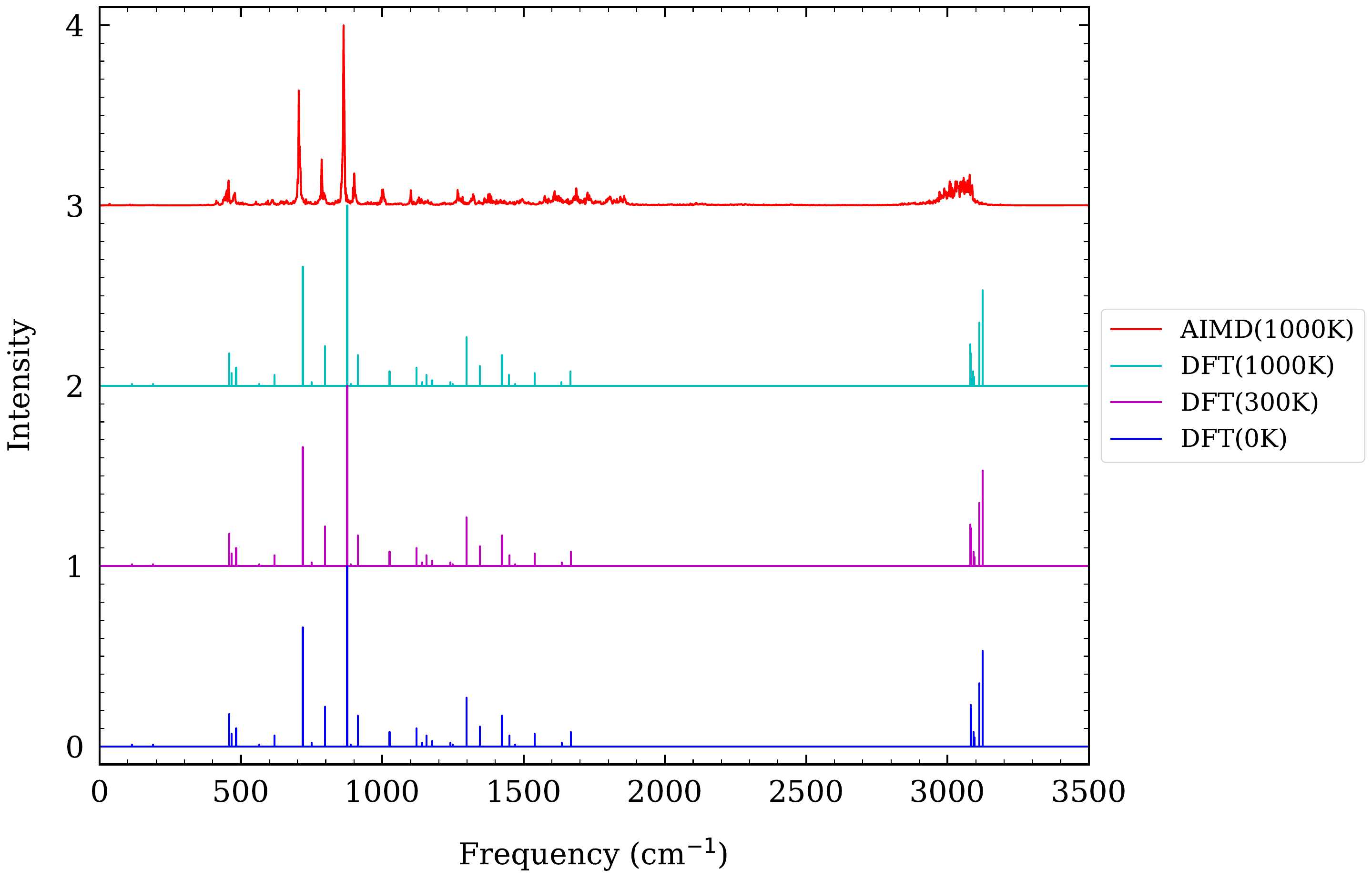} 
\caption{\label{fig:5-acene-NVE-1000K_MD} 
IR spectra of 5-acene in vacuum at the electronic temperature $T_{el}$ = 0 K, 300 K, and 1000 K, computed using FT-TAO-DFT (via NMA). 
For comparison, the IR spectrum obtained with FT-TAO-AIMD simulations at 1000 K is also shown. 
The IR spectra are normalized to have a maximum intensity of one, and are vertically offset from each other by the same value for clarity.} 
\end{figure}

\begin{figure} 
\includegraphics[scale=0.32]{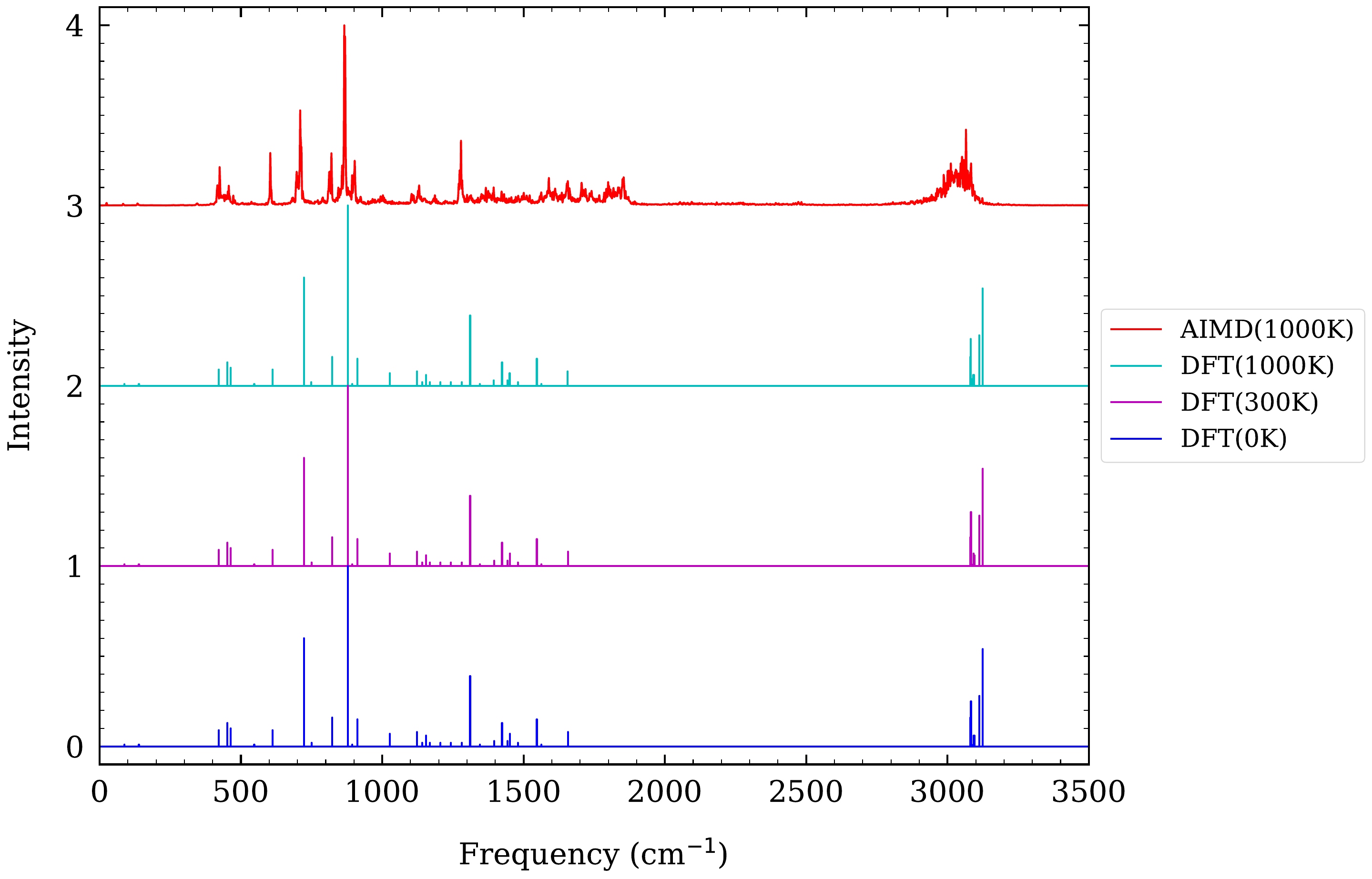} 
\caption{\label{fig:6-acene-NVE-1000K_MD} 
IR spectra of 6-acene in vacuum at the electronic temperature $T_{el}$ = 0 K, 300 K, and 1000 K, computed using FT-TAO-DFT (via NMA). 
For comparison, the IR spectrum obtained with FT-TAO-AIMD simulations at 1000 K is also shown. 
The IR spectra are normalized to have a maximum intensity of one, and are vertically offset from each other by the same value for clarity.} 
\end{figure}

We first examine the electronic temperature effects on the IR spectra of $n$-acenes ($n$ = 2--6), by performing FT-TAO-DFT-based NMA calculations at the electronic temperature $T_{el}$ = 0 K, 300 K, and 1000 K. 
Here, each NMA calculation is performed at the FT-TAO-DFT optimized geometry of $n$-acene at $T_{el}$. As shown 
in \Cref{fig:2-acene-NVE-1000K_MD,fig:3-acene-NVE-1000K_MD,fig:4-acene-NVE-1000K_MD,fig:5-acene-NVE-1000K_MD,fig:6-acene-NVE-1000K_MD}, the NMA-based IR spectra of $n$-acenes are very insensitive to 
electronic temperature changes, suggesting that the electronic temperature effects on the NMA-based IR spectra are rather minor. This is also consistent with the aforementioned finding that the electronic thermal 
ensembles of $n$-acene at $T_{el} \le 1000$ K are mainly contributed by the electronic GS. Therefore, at $T_{el} \le 1000$ K, the NMA-based IR spectra of $n$-acenes obtained with FT-TAO-DFT can be approximated by 
those obtained with TAO-DFT (i.e., FT-TAO-DFT at $T_{el}$ = 0 K) \cite{tao-aimd}. 

Since the NMA-based IR spectra are based on the harmonic approximation, we further investigate the effects of anharmonicity on the IR spectra of $n$-acenes, by performing FT-TAO-AIMD simulations at $T$ = 1000 K 
(with $T_{el}$ = $T$). Along each FT-TAO-AIMD equilibrated trajectory, the IR intensity $I(\omega)$ of $n$-acene can be obtained from the Fourier transform of the autocorrelation function (ACF) of the dipole moment (i.e., 
including both the electronic and nuclear contributions) ${\vec{\mu}}$ of $n$-acene \cite{McQuarrie,tao-aimd,Brehm-2020,Thomas-2013,Thomas-2015,Dutta-Chowdhury-2019}: 
\begin{equation}\label{eq:IR-intensity-1} 
I(\omega) \propto \omega^{2} \int_{-\infty}^{\infty} \langle\vec{\mu}(0) \cdot \vec{\mu}(t) \rangle e^{-i \omega t} dt, 
\end{equation} 
with $\omega$ being the vibrational frequency. Here, a quantum correction factor $\beta \hbar \omega / (1 - e^{-\beta \hbar \omega})$ \cite{Ramirez-2004,Joalland-2010,Thomas-2013} has been taken into account, 
with $\beta = (k_{B} T)^{-1}$. The IR intensity $I(\omega)$ can also be expressed as \cite{Thomas-2013,tao-aimd} 
\begin{equation}\label{eq:IR-intensity-2} 
I(\omega) \propto \int_{-\infty}^{\infty} \langle\dot{\vec{\mu}}(0)\cdot \dot{\vec{\mu}}(t)\rangle e^{-i\omega t} dt, 
\end{equation} 
which is proportional to the Fourier transform of the ACF of $\dot{\vec{\mu}}$ (i.e., the time derivative of the dipole moment) of $n$-acene. In the present study, the IR spectrum of $n$-acene is computed using 
Eq.\ (\ref{eq:IR-intensity-2}). To smooth the IR spectrum, a Gaussian window function $e^{-\frac{\sigma t^{2}}{2 \tau^{2}}}$ \cite{Gaigeot-2005,Gaigeot-2010,Vitale-2015,tao-aimd} has been applied in the time domain to 
reduce the numerical noise, where $\sigma = 10$ and $\tau$ is the total time duration of the FT-TAO-AIMD equilibrated trajectory. The reported IR spectrum of $n$-acene is an average over four different FT-TAO-AIMD 
equilibrated trajectories. 

As presented in \Cref{fig:2-acene-NVE-1000K_MD,fig:3-acene-NVE-1000K_MD,fig:4-acene-NVE-1000K_MD,fig:5-acene-NVE-1000K_MD,fig:6-acene-NVE-1000K_MD}, red-shifts (i.e., shifts towards lower frequencies) are 
observed in the FT-TAO-AIMD-based IR spectra at 1000 K, when compared with the corresponding IR spectra at 300 K (i.e., approximately given by the TAO-AIMD-based IR spectra at 300 K, e.g., see Figures 6 to 10 in the 
TAO-AIMD paper \cite{tao-aimd}). At the frequency $\tilde{\nu}$ = 1000 cm$^{-1}$ or below, the FT-TAO-AIMD-based IR spectra are insensitive to nuclear temperature changes (i.e., negligible red-shifts are observed), and 
the NMA-based IR spectra show strong similarities to the FT-TAO-AIMD-based IR spectra. This implies that the harmonic approximation used in the NMA works well at low frequencies (i.e., below 1000 cm$^{-1}$). However, 
as the frequency increases, the effects of anharmonicity on the IR spectra become increasingly important (e.g., some bands in the range of 1600--2000 cm$^{-1}$ from the FT-TAO-AIMD-based IR spectra are absent in the 
NMA-based IR spectra), and the red-shifts become increasingly pronounced, especially around $\tilde{\nu}$ = 3000 cm$^{-1}$ (i.e., close to one of the UIR emission bands at the wavelength $\lambda$ = 3.3 $\mu m$. 
These findings are consistent with the previous studies of IR spectra of PAHs \cite{PAH-IR-T,PAH-IR}.

\subsection{Acenes in an Ar matrix}\label{acene-ar} 

\subsubsection{Radical nature} 

\begin{figure} 
\includegraphics[scale=0.36]{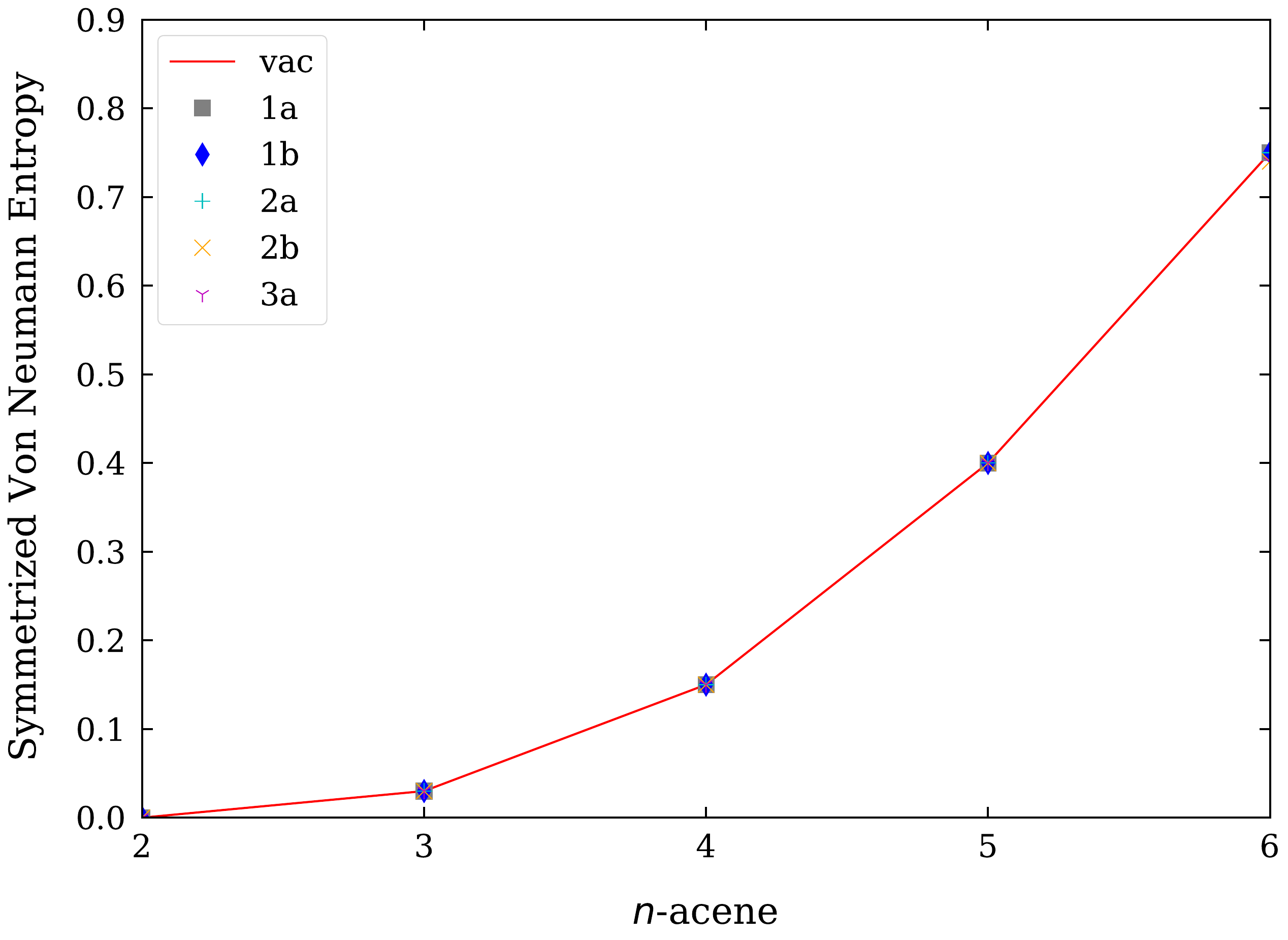} 
\caption{\label{fig:acene-Ar-SVN} 
Symmetrized von Neumann entropy ($S_{\text{vN}}$) of $n$-acene 
inserted into an Ar matrix at various positions (1a, 1b, 2a, 2b, and 3a) at the electronic temperature $T_{el}$ = 0 K, computed using FT-TAO-QM/MM. 
For comparison, the corresponding symmetrized von Neumann entropy in vacuum, computed using FT-TAO-DFT, is also shown.} 
\end{figure}

\begin{figure} 
\includegraphics[scale=0.36]{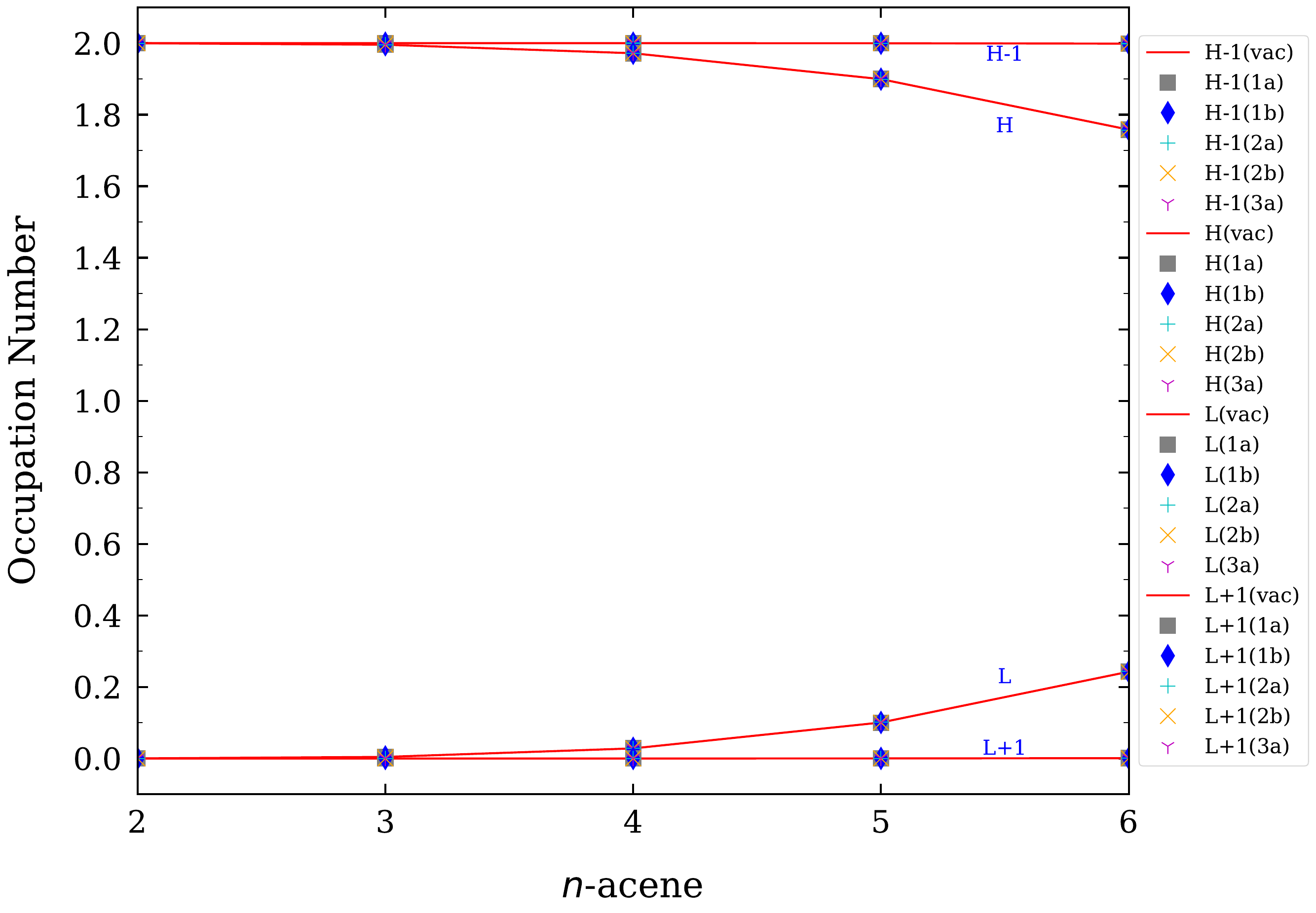} 
\caption{\label{fig:acene-Ar-OCC} 
Active orbital occupation numbers ($f_{\text{H-1}}$, $f_{\text{H}}$, $f_{\text{L}}$, and $f_{\text{L+1}}$) of $n$-acene 
inserted into an Ar matrix at various positions (1a, 1b, 2a, 2b, and 3a) at the electronic temperature $T_{el}$ = 0 K, computed using FT-TAO-QM/MM. 
The HOMO/LUMO is denoted as the H/L for brevity. 
For comparison, the corresponding active orbital occupation numbers in vacuum, computed using FT-TAO-DFT, are also shown.} 
\end{figure}

To investigate the radical nature of $n$-acene in an Ar matrix, FT-TAO-QM/MM calculations are performed for the symmetrized von Neumann entropy $S_{\text{vN}}$ (see \Cref{fig:acene-Ar-SVN}) and active orbital 
occupation numbers (see \Cref{fig:acene-Ar-OCC}) of $n$-acene ($n$ = 2--6) inserted into an Ar matrix at various positions (1a, 1b, 2a, 2b, and 3a, as illustrated in \Cref{fig:position}) at the electronic temperature 
$T_{el}$ = 0 K. Here, each calculation is performed at the FT-TAO-QM/MM optimized geometry of $n$-acene in the Ar matrix at $T_{el}$ = 0 K. 

Relative to the corresponding symmetrized von Neumann entropy $S_{\text{vN}}$ and active orbital occupation numbers in vacuum obtained with FT-TAO-DFT, the Ar matrix has minimal impact on these properties, 
regardless of the position of $n$-acene in the Ar matrix. This can be attributed to the weak non-covalent interactions between $n$-acene and Ar atoms, yielding the minimally distorted $n$-acene geometry, and hence, 
the essentially unchanged radical nature of $n$-acene during the co-deposition. Therefore, similar to the vacuum counterparts at $T_{el}$ = 0 K, the smaller $n$-acenes ($n$ = 2--5) in the Ar matrix should possess 
non-radical nature, and 6-acene in the Ar matrix should possess noticeable di-radical nature, regardless of the position of $n$-acene in the Ar matrix. This well justifies the use of FT-TAO-QM/MM in the present study, 
due to the radical nature of 6-acene and a large number of Ar atoms involved.

\subsubsection{IR spectra} 

\begin{figure} 
\includegraphics[scale=0.36]{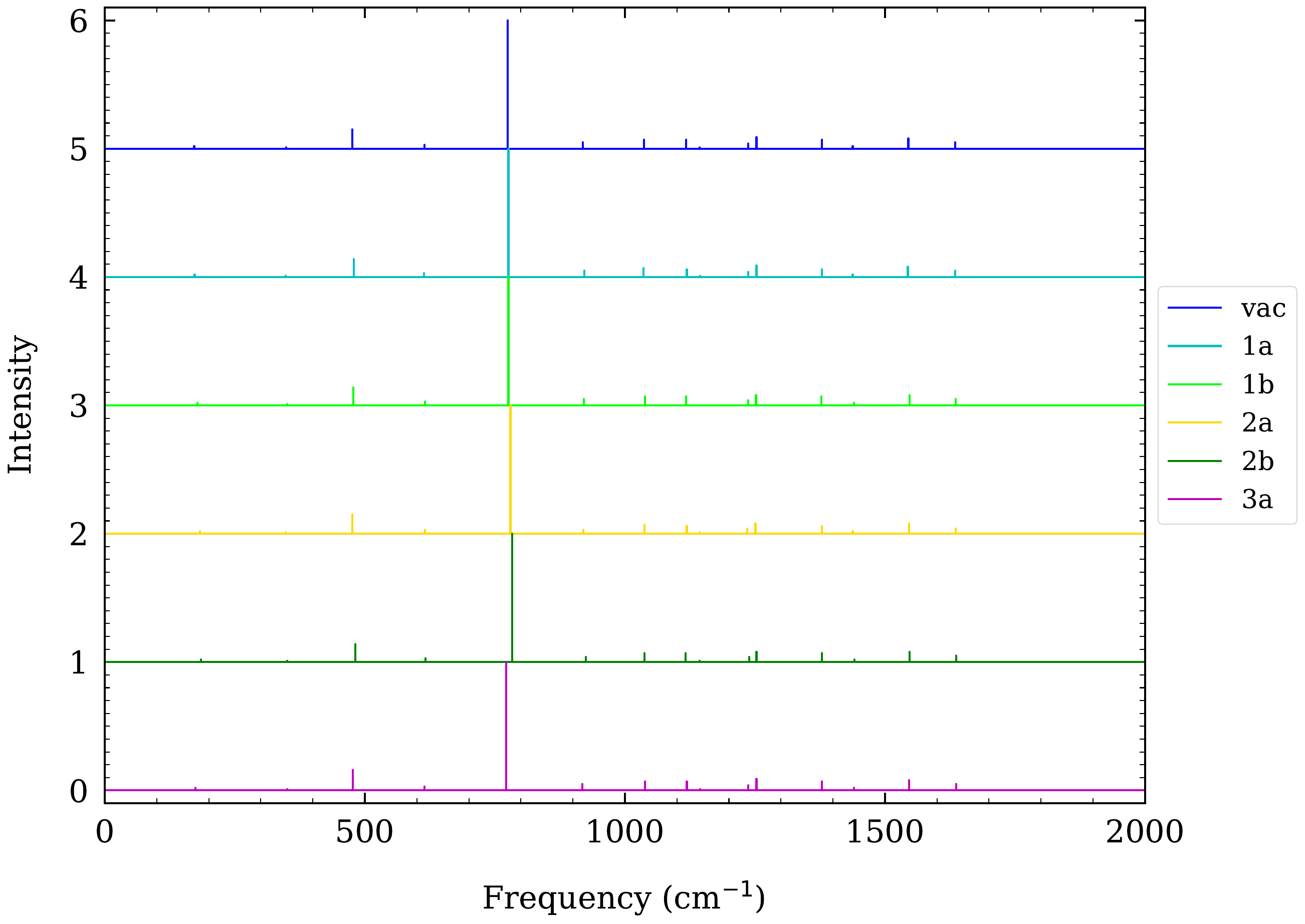} 
\caption{\label{fig:2_acene-NMA-0K-ind} 
IR spectra of 2-acene 
inserted into an Ar matrix at various positions (1a, 1b, 2a, 2b, and 3a) at the electronic temperature $T_{el}$ = 0 K, computed using FT-TAO-QM/MM (via NMA). 
For comparison, the IR spectrum in vacuum, computed using FT-TAO-DFT (via NMA), is also shown. 
The IR spectra are normalized to have a maximum intensity of one, and are vertically offset from each other by the same value for clarity.} 
\end{figure}

\begin{figure} 
\includegraphics[scale=0.36]{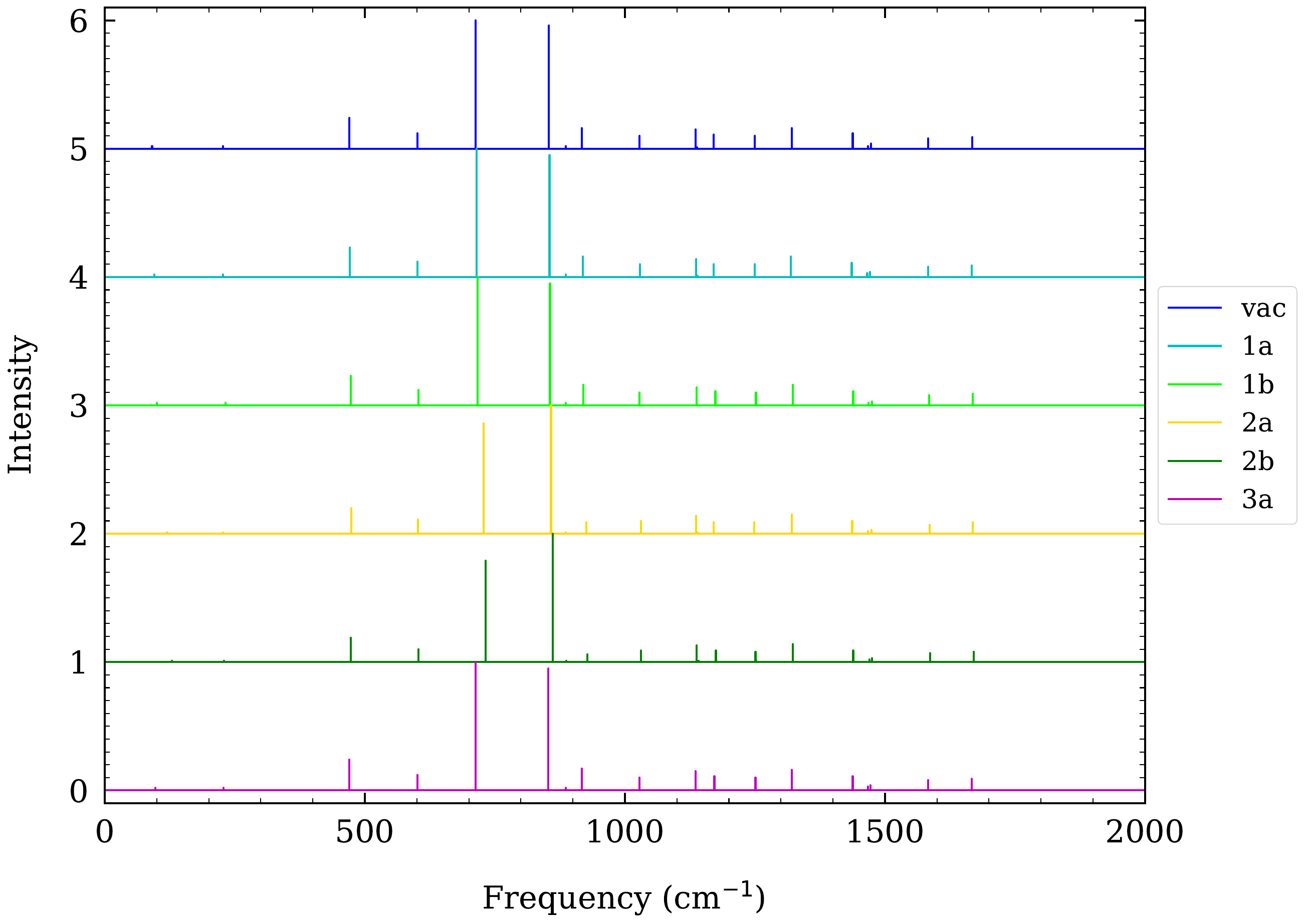} 
\caption{\label{fig:3_acene-NMA-0K-ind} 
IR spectra of 3-acene 
inserted into an Ar matrix at various positions (1a, 1b, 2a, 2b, and 3a) at the electronic temperature $T_{el}$ = 0 K, computed using FT-TAO-QM/MM (via NMA). 
For comparison, the IR spectrum in vacuum, computed using FT-TAO-DFT (via NMA), is also shown. 
The IR spectra are normalized to have a maximum intensity of one, and are vertically offset from each other by the same value for clarity.} 
\end{figure}

\begin{figure} 
\includegraphics[scale=0.36]{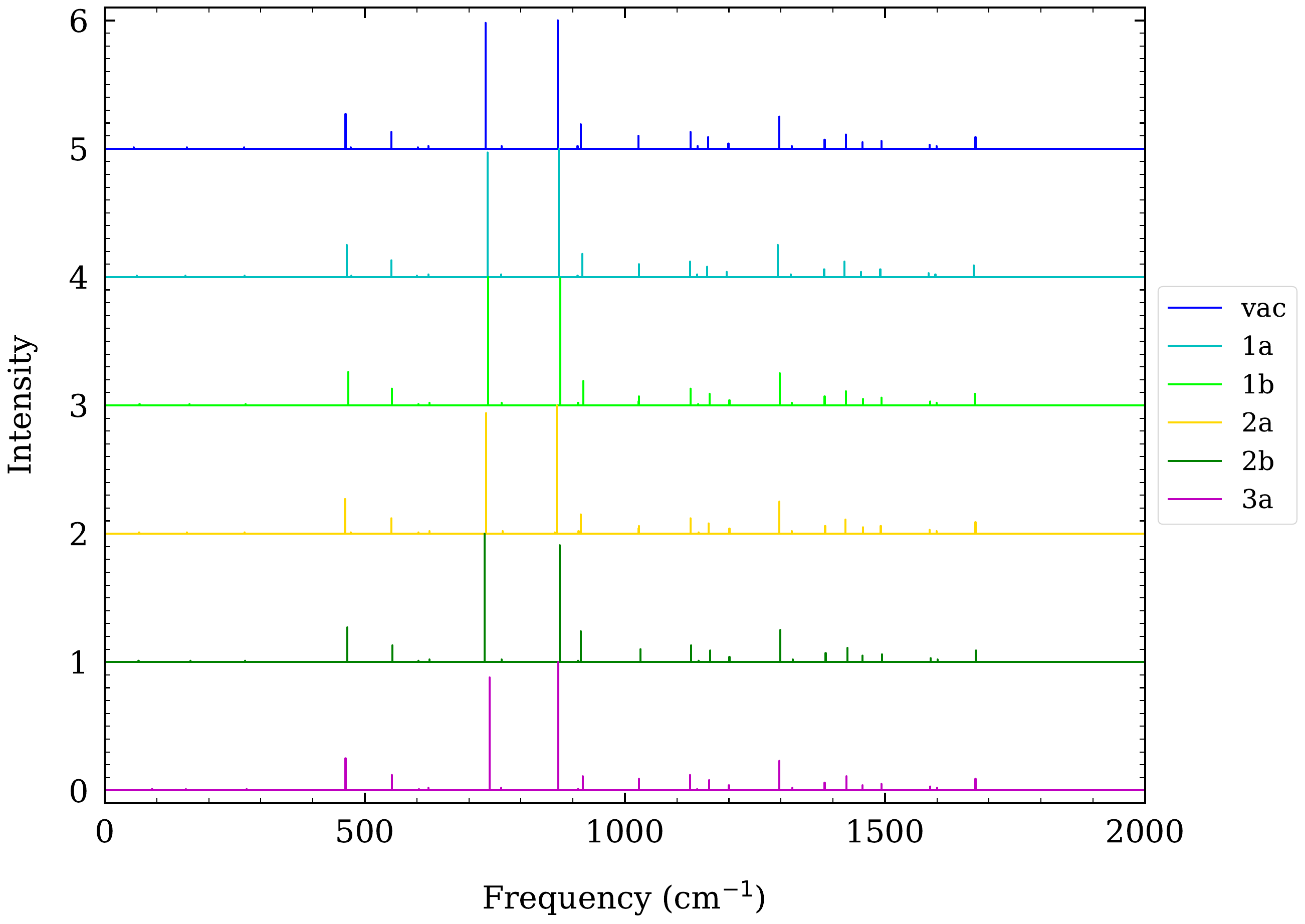} 
\caption{\label{fig:4_acene-NMA-0K-ind} 
IR spectra of 4-acene 
inserted into an Ar matrix at various positions (1a, 1b, 2a, 2b, and 3a) at the electronic temperature $T_{el}$ = 0 K, computed using FT-TAO-QM/MM (via NMA). 
For comparison, the IR spectrum in vacuum, computed using FT-TAO-DFT (via NMA), is also shown. 
The IR spectra are normalized to have a maximum intensity of one, and are vertically offset from each other by the same value for clarity.} 
\end{figure}

\begin{figure} 
\includegraphics[scale=0.36]{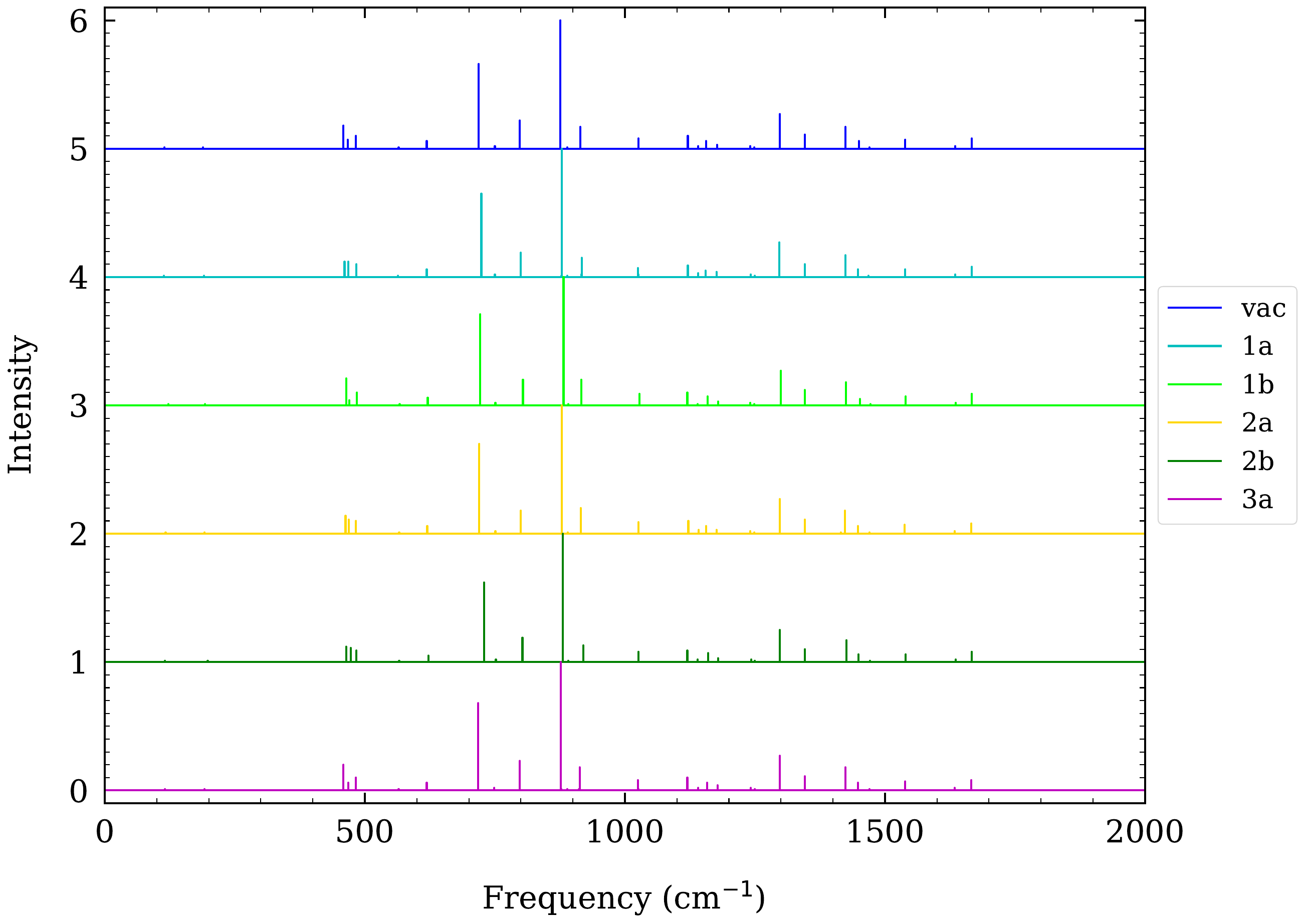} 
\caption{\label{fig:5_acene-NMA-0K-ind} 
IR spectra of 5-acene 
inserted into an Ar matrix at various positions (1a, 1b, 2a, 2b, and 3a) at the electronic temperature $T_{el}$ = 0 K, computed using FT-TAO-QM/MM (via NMA). 
For comparison, the IR spectrum in vacuum, computed using FT-TAO-DFT (via NMA), is also shown. 
The IR spectra are normalized to have a maximum intensity of one, and are vertically offset from each other by the same value for clarity.} 
\end{figure}

\begin{figure} 
\includegraphics[scale=0.36]{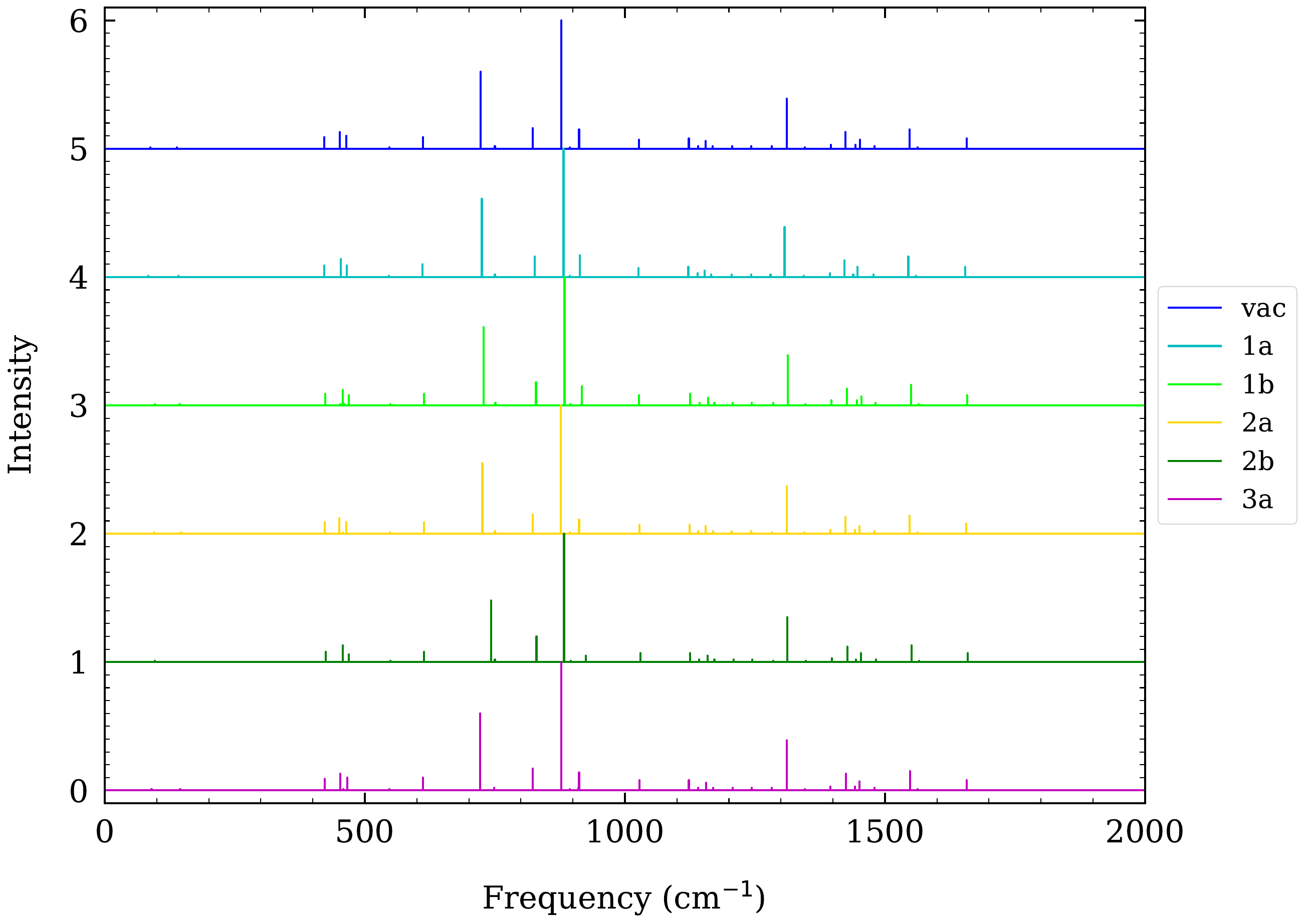} 
\caption{\label{fig:6_acene-NMA-0K-ind} 
IR spectra of 6-acene 
inserted into an Ar matrix at various positions (1a, 1b, 2a, 2b, and 3a) at the electronic temperature $T_{el}$ = 0 K, computed using FT-TAO-QM/MM (via NMA). 
For comparison, the IR spectrum in vacuum, computed using FT-TAO-DFT (via NMA), is also shown. 
The IR spectra are normalized to have a maximum intensity of one, and are vertically offset from each other by the same value for clarity.} 
\end{figure}

As presented in \Cref{fig:2_acene-NMA-0K-ind,fig:3_acene-NMA-0K-ind,fig:4_acene-NMA-0K-ind,fig:5_acene-NMA-0K-ind,fig:6_acene-NMA-0K-ind}, the IR spectra of $n$-acene inserted into an Ar matrix at various 
positions (1a, 1b, 2a, 2b, and 3a) at $T_{el}$ = 0 K are computed using FT-TAO-QM/MM-based NMA (also see the SI for detailed results). Here, each NMA calculation is performed at the FT-TAO-QM/MM optimized 
geometry of $n$-acene in the Ar matrix at $T_{el}$ = 0 K. For comparison, the corresponding IR spectra in vacuum at $T_{el}$ = 0 K, computed using FT-TAO-DFT-based NMA, are also provided. 

Owing to the weak non-covalent interactions between $n$-acene and Ar atoms, matrix-induced frequency shifts (i.e., matrix shifts) \cite{matrix-isolation-book,matrix-shift,argon-oniom} 
\begin{equation}\label{eq:matrixs} 
\Delta\tilde{\nu} \equiv \tilde{\nu}^{'} - \tilde{\nu} 
\end{equation} 
can occur in the IR spectra of $n$-acene in the Ar matrix, when compared with the corresponding IR spectra in vacuum. Here, $\tilde{\nu}^{'}$ and $\tilde{\nu}$ are the peak frequencies in the IR spectra of $n$-acene in 
the Ar matrix and in vacuum, respectively. Since the effects of anharmonicity, which are not captured in the NMA, cannot be completely neglected at higher frequencies, the matrix shifts are reported only for the IR bands 
below 2000 cm$^{-1}$. 

The matrix shifts $\Delta\tilde{\nu}$ associated with the peaks of relative intensities $I \ge 0.05$ and whose magnitudes $|\Delta\tilde{\nu}| \ge 10$ cm$^{-1}$ are observed in the IR spectra of 
3-acene at position 2a ($\Delta\tilde{\nu}$ = 15 cm$^{-1}$ at $\tilde{\nu}$ = 713 cm$^{-1}$) and at position 2b ($\Delta\tilde{\nu}$ = 19 cm$^{-1}$ at $\tilde{\nu}$ = 713 cm$^{-1}$ and 
$\Delta\tilde{\nu}$ = 11 cm$^{-1}$ at $\tilde{\nu}$ = 917 cm$^{-1}$), 5-acene at position 2b ($\Delta\tilde{\nu}$ = 10 cm$^{-1}$ at $\tilde{\nu}$ = 719 cm$^{-1}$), and 
6-acene at position 2b ($\Delta\tilde{\nu}$ = 20 cm$^{-1}$ at $\tilde{\nu}$ = 723 cm$^{-1}$ and $\Delta\tilde{\nu}$ = 13 cm$^{-1}$ at $\tilde{\nu}$ = 912 cm$^{-1}$). On the other hand, very minor matrix effects (e.g., with 
$|\Delta\tilde{\nu}| < 2$ cm$^{-1}$) are also observed in the IR spectra of 2-acene at position 1b, 3-acene at positions 1a and 3a, 4-acene at position 2a, 5-acene at position 3a, and 6-acene at position 3a. Therefore, the 
matrix shifts $\Delta\tilde{\nu}$ can greatly depend on the position of $n$-acene in the Ar matrix. This also suggests that the co-deposition procedure of $n$-acene and Ar atoms may affect the IR spectrum of $n$-acene. 
Accordingly, to obtain the more realistic IR spectra of $n$-acene in an Ar matrix, it can be essential to perform FT-TAO-QM/MM-based molecular dynamics (FT-TAO-QM/MM-MD) simulations at extremely low temperatures 
($\approx$ 10 K) \cite{acene-EXP-IR-1,acene-EXP-IR-2}. However, this requires additional techniques to improve the computational efficiency, which we leave for future work.

\section{Conclusions}\label{conclusion} 

In conclusion, we have developed FT-TAO-DFT and related extensions (i.e., FT-TAO-AIMD and FT-TAO-QM/MM) to explore the properties of large MR systems at finite temperatures. The fictitious temperature $\theta$ in 
FT-TAO-DFT can be properly selected to improve the TE density representability for a physical system at the electronic temperature $\theta_{el}$. Consequently, the optimal fictitious temperature $\theta$ in FT-TAO-DFT 
should be both system-dependent and $\theta_{el}$-dependent. By contrast, for FT-KS-DFT \cite{KS-DFT,Mermin} (i.e., FT-TAO-DFT with $\theta = \theta_{el}$), the constraint $\theta = \theta_{el}$ seems unnecessary, and 
can, in fact, limit the TE density representability. 

FT-TAO-DFT is suitable for studying the TE properties of large MR systems at finite electronic temperatures ($\theta_{el} = k_{B} T_{el} \ge 0$). At zero electronic temperature ($T_{el} = 0\ \text{K}$), FT-TAO-DFT reduces 
to TAO-DFT \cite{tao1}. Because of its computational efficiency, FT-TAO-AIMD is promising for exploring the dynamical information of large MR systems at finite temperatures (with the nuclear temperature $T$ = $T_{el}$). 
At very low electronic temperatures ($T_{el} \approx 0\ \text{K}$), FT-TAO-AIMD can be approximated by TAO-AIMD \cite{tao-aimd}. In addition, FT-TAO-QM/MM provides a cost-effective description of the TE properties of 
a QM subsystem with MR character embedded in an MM environment at finite temperatures. 

To highlight their capacities, the FT-TAO-DFT and FT-TAO-AIMD methods have been applied to investigate the radical nature and IR spectra of $n$-acene ($n$ = 2--6) at finite temperatures. The electronic temperature 
effects on the radical nature and IR spectra of $n$-acene at $T_{el} \le 1000$ K are rather minor, suggesting that the electronic thermal ensembles of $n$-acene at $T_{el} \le 1000$ K are mainly contributed by the electronic 
GS (i.e., the electronic thermal ensemble at $T_{el} = 0\ \text{K}$). By contrast, the nuclear temperature effects (i.e., directly related to the nuclear kinetic energy) are shown to be responsible for the change in the radical 
nature and IR spectra of $n$-acene. Besides, the FT-TAO-QM/MM method has been applied to explore the radical nature and IR spectra of $n$-acene ($n$ = 2--6) inserted into an Ar matrix at various positions at 
$T_{el}$ = 0 K. The Ar matrix has minimal impact on the radical nature of $n$-acene (i.e., regardless of the position of $n$-acene in the Ar matrix), while the co-deposition procedure of $n$-acene and Ar atoms may affect 
the IR spectrum of $n$-acene. To obtain the more realistic IR spectra of $n$-acene in an Ar matrix, it can be essential to perform FT-TAO-QM/MM-MD simulations at extremely low temperatures ($\approx$ 10 K). 

Note, however, that there are a few limitations in this work. While the optimal fictitious temperature $\theta$ in FT-TAO-DFT should be both system-dependent and $\theta_{el}$-dependent, in this work, we adopt the optimal 
system-independent and $\theta_{el}$-independent $\theta$ (i.e., 7 mhartree for FT-TAO-LDA) \cite{tao1}, which should work reasonably well for many systems at very low electronic temperatures ($\theta_{el} \approx 0$) 
by construction \cite{tao1,theta2022,tao-rsh}. However, this $\theta$ can be inappropriate for some cases. For example, even at $\theta_{el} = 0$, this $\theta$ can be too large for some SR systems (e.g., the dissociation of 
H$_{2}^{+}$ and He$_{2}^{+}$), while it can be too small for some MR systems (e.g., the dissociation of H$_{2}$ and N$_{2}$) \cite{tao1,spin-symm,tao-rsh}. Besides, for a physical system at considerably high electronic 
temperatures (e.g., warm dense matter (WDM) \cite{WDM1,WDM2,WDM3}), wherein the electronic thermal ensembles include the electronic GS and higher-energy excited states, it remains challenging to determine the 
optimal $\theta$ in FT-TAO-DFT. In general, it is essential to develop a scheme in FT-TAO-DFT to reliably determine the optimal $\theta$ of a physical system at any given electronic temperature $\theta_{el}$. We plan to 
investigate along these lines, and results may be reported elsewhere.

\begin{acknowledgments} 

This work was supported by National Science and Technology Council of Taiwan (Grant Nos.: NSTC114-2112-M-002-033; NSTC113-2112-M-002-032), National Taiwan University, and 
National Center for Theoretical Sciences of Taiwan. We thank Dr. Jung-Hsin Lin and Alvin Hew for useful discussions about QM/MM simulations. 

\end{acknowledgments}

\clearpage 

\clearpage 

\clearpage 

\clearpage 

\clearpage 

\clearpage 

\clearpage 

\clearpage 

\clearpage 

\clearpage 

\clearpage 

\clearpage 

\clearpage 

\clearpage 

\clearpage 

\clearpage 

\clearpage 

\clearpage 

\clearpage 
\begin{table} 
\caption{\label{table:LJ-coeff} 
Atomic Lennard-Jones parameters, $\epsilon$ (in kcal/mol) and $r_{\text{min}} \equiv 2^{1/6} \sigma$ (in {\AA}), adopted in the FT-TAO-QM/MM calculations. 
The parameters are taken from the OPLS-AA force field \cite{OPLSAA-1,OPLSAA-2}.} 
\begin{tabular}{|c|c|c|} 
\hline 
atom & $\epsilon$ & $r_{\text{min}}$  \\ 
\hline 
H & 0.0300 & 2.4200 \\ 
\hline 
C & 0.0700 & 3.5500 \\ 
\hline 
Ar & 0.2339 & 3.4010 \\ 
\hline 
\end{tabular} 
\end{table} 


\newpage 
\begin{thebibliography}{99} 
\expandafter\ifx\csname url\endcsname\relax 
\def\url#1{\texttt{#1}}\fi 
\expandafter\ifx\csname urlprefix\endcsname\relax\def\urlprefix{URL }\fi
\providecommand{\bibinfo}[2]{#2}
\providecommand{\eprint}[2][]{\url{#2}}


\bibitem{HK} 
P. Hohenberg and W. Kohn, 
Phys. Rev. {\bf 136}, B864 (1964). 

\bibitem{KS-DFT} 
W. Kohn and L. J. Sham, 
Phys. Rev. {\bf 140}, A1133 (1965). 

\bibitem{parr-yang} 
R. G. Parr and W. Yang, 
{\it Density-Functional Theory of Atoms and Molecules} 
(Oxford University, New York, 1989). 

\bibitem{Kohanoff} 
J. Kohanoff, 
{\it Electronic Structure Calculations for Solids and Molecules: Theory and Computational Methods} 
(Cambridge University, New York, 2006). 

\bibitem{Jensen} 
F. Jensen, 
{\it Introduction to Computational Chemistry} 
(Wiley, New York, 2007). 

\bibitem{DFTReview} 
S. K\"{u}mmel and L. Kronik, 
Rev. Mod. Phys. {\bf 80}, 3 (2008). 

\bibitem{Yang12} 
A. J. Cohen, P. Mori-S\'{a}nchez, and W. Yang, 
Chem. Rev. {\bf 112}, 289 (2012). 

\bibitem{Mermin} 
N. D. Mermin, 
Phys. Rev. {\bf 137}, A1441 (1965). 

\bibitem{gross-2011} 
S. Pittalis, C. R. Proetto, A. Floris, A. Sanna, C. Bersier, K. Burke, and E. K. U. Gross, 
Phys. Rev. Lett. {\bf 107}, 163001 (2011). 

\bibitem{LDAX} 
P. A. M. Dirac, 
Proc. Cambridge Philos. Soc. {\bf 26}, 376 (1930). 

\bibitem{LDAC} 
J. P. Perdew and Y. Wang, 
Phys. Rev. B {\bf 45}, 13244 (1992). 

\bibitem{B88} 
A. D. Becke, 
Phys. Rev. A {\bf 38}, 3098 (1988). 

\bibitem{LYP} 
C. Lee, W. Yang, and R. G. Parr, 
Phys. Rev. B {\bf 37}, 785 (1988). 

\bibitem{PBE} 
J. P. Perdew, K. Burke, and M. Ernzerhof, 
Phys. Rev. Lett. {\bf 77}, 3865 (1996). 

\bibitem{hybrid1} 
A. D. Becke, 
J. Chem. Phys. {\bf 98}, 1372 (1993). 

\bibitem{hybrid2} 
A. D. Becke, 
J. Chem. Phys. {\bf 98}, 5648 (1993). 

\bibitem{hybrid0} 
J. P. Perdew, M. Ernzerhof, and K. Burke, 
J. Chem. Phys. {\bf 105}, 9982 (1996). 

\bibitem{LC-DFT} 
A. Savin, 
in {\it Recent Developments and Applications of Modern Density Functional Theory}, 
edited by J. M. Seminario (Elsevier, Amsterdam, 1996), pp. 327--357. 

\bibitem{LCHirao} 
H. Iikura, T. Tsuneda, T. Yanai, and K. Hirao, 
J. Chem. Phys. {\bf 115}, 3540 (2001). 

\bibitem{wB97X} 
J.-D. Chai and M. Head-Gordon, 
J. Chem. Phys. {\bf 128}, 084106 (2008). 

\bibitem{Yang-science-2008} 
A. J. Cohen, P. Mori-S\'{a}nchez, and W. Yang, 
Science {\bf 321}, 792 (2008). 

\bibitem{Yang-JCP-2008} 
A. J. Cohen, P. Mori-S\'{a}nchez, and W. Yang, 
J. Chem. Phys. {\bf 129}, 121104 (2008). 

\bibitem{ftlda} 
V. V. Karasiev, T. Sjostrom, J. Dufty, and S. B. Trickey, 
Phys. Rev. Lett. {\bf 112}, 076403 (2014). 

\bibitem{ftgga} 
V. V. Karasiev, J. W. Dufty, and S. B. Trickey, 
Phys. Rev. Lett. {\bf 120}, 076401 (2018). 

\bibitem{ftgh} 
D. I. Mihaylov, V. V. Karasiev, and S. X. Hu, 
Phys. Rev. B {\bf 101}, 245141 (2020). 

\bibitem{ftrsh} 
A. A. Ellaboudy, V. V. Karasiev, D. I. Mihaylov, K. P. Hilleke, and S. X. Hu, 
Phys. Rev. B {\bf 112}, 155154 (2025). 

\bibitem{CASSCF} 
B. O. Roos, P. R. Taylor, and P. E. M. Siegbahn, 
Chemical Physics {\bf 48}, 157 (1978). 

\bibitem{CASPT2} 
K. Andersson, P.-Å. Malmqvist, and B. O. Roos, 
J. Chem. Phys. {\bf 96}, 1218 (1992). 

\bibitem{dmrg} 
J. Hachmann, J. J. Dorando, M. Avil{'e}s, and G. K.-L. Chan, 
J. Chem. Phys. {\bf 127}, 134309 (2007). 

\bibitem{2-RDMa} 
G. Gidofalvi and D. A. Mazziotti, 
J. Chem. Phys. {\bf 129}, 134108 (2008). 

\bibitem{cote2015} 
G. Gryn'ova, M. L. Coote, and C. Corminboeuf, 
Wiley Interdiscip. Rev. Comput. Mol. Sci. {\bf 5}, 440 (2015). 

\bibitem{tao1} 
J.-D. Chai, 
J. Chem. Phys. {\bf 136}, 154104 (2012). 

\bibitem{Lowdin-1955} 
P.-O. L\"{o}wdin, 
Phys. Rev. {\bf 97}, 1474 (1955). 

\bibitem{Lowdin-1956} 
P.-O. L\"{o}wdin and H. Shull, 
Phys. Rev. {\bf 101}, 1730 (1956). 

\bibitem{spin-symm} 
Y.-Y. Wang and J.-D. Chai, 
Phys. Rev. A {\bf 109}, 062808 (2024). 

\bibitem{tao2} 
J.-D. Chai, 
J. Chem. Phys. {\bf 140}, 18A521 (2014). 

\bibitem{tao3} 
J.-D. Chai, 
J. Chem. Phys. {\bf 146}, 044102 (2017). 

\bibitem{tao-rsh} 
S. Li and J.-D. Chai, 
J. Chem. Theory Comput. {\bf 21}, 9538 (2025). 

\bibitem{theta2022} 
B.-J. Chen and J.-D. Chai, 
RSC Adv. {\bf 12}, 12193 (2022). 

\bibitem{tao4} 
C.-Y. Lin, K. Hui, J.-H. Chung, and J.-D. Chai, 
RSC Adv. {\bf 7}, 50496 (2017). 

\bibitem{tao-gnr} 
C.-S. Wu and J.-D. Chai, 
J. Chem. Theory Comput. {\bf 11}, 2003 (2015). 

\bibitem{NK} 
C.-N. Yeh and J.-D. Chai, 
Sci. Rep. {\bf 6}, 30562 (2016). 

\bibitem{H2S1} 
S. Seenithurai and J.-D. Chai, 
Sci. Rep. {\bf 6}, 33081 (2016). 

\bibitem{cycl} 
C.-S. Wu, P.-Y. Lee, and J.-D. Chai, 
Sci. Rep. {\bf 6}, 37249 (2016). 

\bibitem{coronene} 
C.-N. Yeh, C. Wu, H. Su, and J.-D. Chai, 
RSC Adv. {\bf 8}, 34350 (2018). 

\bibitem{mobius} 
J.-H. Chung and J.-D. Chai, 
Sci. Rep. {\bf 9}, 2907 (2019). 

\bibitem{BNR} 
S. Seenithurai and J.-D. Chai, 
Sci. Rep. {\bf 9}, 12139 (2019). 

\bibitem{CC} 
S. Seenithurai and J.-D. Chai, 
Sci. Rep. {\bf 10}, 13133 (2020). 

\bibitem{Manassir2020} 
M. Manassir and A. H. Pakiari, 
J. Mol. Graph. Model. {\bf 99}, 107643 (2020). 

\bibitem{HansonHeine2020} 
M. W. D. Hanson-Heine, 
Chem. Phys. Lett. {\bf 739}, 137012 (2020). 

\bibitem{HansonHeine2020b} 
M. W. D. Hanson-Heine, D. M. Rogers, S. Woodward, and J. D. Hirst, 
J. Phys. Chem. Lett. {\bf 11}, 3769 (2020). 

\bibitem{HansonHeine2020c} 
M. W. D. Hanson-Heine and J. D. Hirst, 
J. Phys. Chem. A {\bf 124}, 5408 (2020). 

\bibitem{SS21} 
S. Seenithurai and J.-D. Chai, 
Nanomaterials {\bf 11}, 2224 (2021). 

\bibitem{Bettinger2021} 
C. T\"{o}nshoff and H. F. Bettinger, 
Chem. Eur. J. {\bf 27}, 3193 (2021). 

\bibitem{Bettinger2021a} 
D. Gupta, A. Omont, and H. F. Bettinger, 
Chem. Eur. J. {\bf 27}, 4605 (2021). 

\bibitem{CCC} 
C.-C. Chen and J.-D. Chai, 
Nanomaterials {\bf 12}, 3943 (2022). 

\bibitem{tao-vib2} 
M. W. D. Hanson-Heine, 
J. Phys. Chem. A {\bf 126}, 7273 (2022). 

\bibitem{Nieman2023} 
R. Nieman, J. R. Carvalho, B. Jayee, A. Hansen, A. J. Aquino, M. Kertesz, and H. Lischka, 
Phys. Chem. Chem. Phys. {\bf 25}, 27380 (2023). 

\bibitem{SS24} 
S. Seenithurai and J.-D. Chai, 
Molecules {\bf 29}, 349 (2024). 

\bibitem{Efields} 
C.-Y. Chen and J.-D. Chai, 
Molecules {\bf 29}, 4245 (2024). 

\bibitem{Bettinger2024} 
A. Somani, D. Gupta, and H. F. Bettinger, 
J. Phys. Chem. A {\bf 128}, 6847 (2024). 

\bibitem{Bettinger2025} 
A. Somani, D. Gupta, and H. F. Bettinger, 
Chemistry {\bf 7}, 62 (2025). 

\bibitem{HansonHeine2025} 
M. W. D. Hanson-Heine, 
J. Phys. Chem. A {\bf 129}, 8601 (2025). 

\bibitem{tao-aimd} 
S. Li and J.-D. Chai, 
Front. Chem. {\bf 8}, 589432 (2020). 

\bibitem{TAO-PCM} 
S. Seenithurai and J.-D. Chai, 
Nanomaterials {\bf 13}, 1593 (2023). 

\bibitem{rttao} 
H.-Y. Tsai and J.-D. Chai, 
Molecules {\bf 28}, 7247 (2023). 

\bibitem{AIMD2002} 
M. E. Tuckerman, 
J. Phys.: Condens. Matter {\bf 14}, R1297 (2002). 

\bibitem{QMMM-1} 
A. Warshel and M. Levitt, 
J. Mol. Biol. {\bf 103}, 227 (1976). 

\bibitem{QMMM-2} 
M. J. Field, P. A. Bash, and M. Karplus, 
J. Comput. Chem. {\bf 11}, 700 (1990). 

\bibitem{QMMM-3} 
P. D. Lyne, M. Hodoscek, and M. Karplus, 
J. Phys. Chem. A {\bf 103}, 3462 (1999). 

\bibitem{Mermin-SP-0} 
A. K. Rajagopal, 
Adv. Chem. Phys. {\bf 41}, 59 (1980). 

\bibitem{Mermin-SP-1} 
R. Balawender and A. Holas, 
arXiv:0901.1060 (2009). 

\bibitem{Mermin-SP-2} 
R. Balawender and A. Holas, 
arXiv:0904.3990 (2009). 

\bibitem{perrot-1979} 
F. Perrot, 
Phys. Rev. A {\bf 20}, 586 (1979). 

\bibitem{Karasiev-2012} 
V. V. Karasiev, T. Sjostrom, and S. B. Trickey, 
Phys. Rev. B {\bf 86}, 115101 (2012). 

\bibitem{ft-rdmt} 
T. Baldsiefen, A. Cangi, and E. K. U. Gross, 
Phys. Rev. A {\bf 92}, 052514 (2015). 

\bibitem{free-energy-surface-2} 
J. L. Alonso, A. Castro, J. Clemente-Gallardo, P. Echenique, J. J. Mazo, V. Polo, A. Rubio, and D. Zueco, 
J. Chem. Phys. {\bf 137}, 22A533 (2012). 

\bibitem{McQuarrie} 
D. A. McQuarrie, 
{\it Statistical Mechanics} 
(Harper \& Row, NewYork, 1976). 

\bibitem{BO-approx} 
M. Born and R. Oppenheimer, 
Ann. Phys. {\bf 389}, 457 (1927). 

\bibitem{qmmm-review} 
U. N. Morzan, D. J. Alonso de Armiño, N. O. Foglia, F. Ramírez, M. C. González Lebrero, D. A. Scherlis, and D. A. Estrin, 
Chem. Rev. {\bf 118}, 4071 (2018). 

\bibitem{matrix-isolation-1} 
E. Whittle, D. A. Dows, and G. C. Pimentel, 
J. Chem. Phys. {\bf 22}, 1943 (1954). 

\bibitem{matrix-isolation-2} 
I. Norman and G. Porter, 
Nature {\bf 174}, 508 (1954). 

\bibitem{matrix-isolation-book} 
S. Cradock and A. Hinchcliffe, 
{\it Matrix Isolation: A Technique for the Study of Reactive Inorganic Species} 
(Cambridge University, New York, 1975). 

\bibitem{acene-EXP-IR-1} 
D. M. Hudgins and S. A. Sandford, 
J. Phys. Chem. A {\bf 102}, 329 (1998). 

\bibitem{acene-EXP-IR-2} 
D. M. Hudgins and L. J. Allamandola, 
J. Phys. Chem. {\bf 99}, 3033 (1995). 

\bibitem{Q-Chem5} 
E. Epifanovsky \textit{et al.}, 
J. Chem. Phys. {\bf 155}, 084801 (2021). 

\bibitem{NH-chain}
G. J. Martyna, M. L. Klein, and M. Tuckerman, 
J. Chem. Phys. {\bf 97}, 2635 (1992). 

\bibitem{thermostat-infer} 
M.-P. Gaigeot and M. Sprik, 
J. Phys. Chem. B {\bf 107}, 10344 (2003). 

\bibitem{Brehm-2020} 
M. Brehm, M. Thomas, S. Gehrke, and B. Kirchner, 
J. Chem. Phys. {\bf 152}, 164105 (2020). 

\bibitem{Thomas-2013} 
M. Thomas, M. Brehm, R. Fligg, P. Vöhringer, and B. Kirchner, 
Phys. Chem. Chem. Phys. {\bf 15}, 6608 (2013). 

\bibitem{Thomas-2015} 
M. Thomas, M. Brehm, and B. Kirchner, 
Phys. Chem. Chem. Phys. {\bf 17}, 3207 (2015). 

\bibitem{Brehm-Kirchner-2011} 
M. Brehm and B. Kirchner, 
J. Chem. Inf. Model. {\bf 51}, 2007 (2011). 

\bibitem{argon-exp} 
O. G. Peterson, D. N. Batchelder, and R. O. Simmons, 
Phys. Rev. {\bf 150}, 703 (1966). 

\bibitem{OPLSAA-1}
G. A. Kaminski, R. A. Friesner, J. Tirado-Rives, and W. L. Jorgensen, 
J. Phys. Chem. B {\bf 105}, 6474 (2001). 

\bibitem{OPLSAA-2} 
W. L. Jorgensen, D. S. Maxwell, and J. Tirado-Rives, 
J. Am. Chem. Soc. {\bf 118}, 11225 (1996). 

\bibitem{insert} 
W. Sander, S. Roy, I. Polyak, J. M. Ramirez-Anguita, and E. Sanchez-Garcia, 
J. Am. Chem. Soc. {\bf 134}, 8222 (2012). 

\bibitem{In61} 
D. C. Harris and M. D. Bertolucci, 
{\it Symmetry and Spectroscopy: An Introduction to Vibrational and Electronic Spectroscopy} 
(Oxford University, New York, 1978). 

\bibitem{NMA} 
E. B. Wilson Jr., J. C. Decius, and P. C. Cross, 
{\it Molecular Vibrations: The Theory of Infrared and Raman Vibrational Spectra} 
(Dover, New York, 1980). 

\bibitem{In62} 
M.-P. Gaigeot, M. Martinez, and R. Vuilleumier, 
Mol. Phys. {\bf 105}, 2857 (2007). 

\bibitem{PHVA} 
H. Li and J. H. Jensen, 
Theor. Chem. Acc. {\bf 107}, 211 (2002). 

\bibitem{PHVA-1} 
M. A. Mroginski, 
in {\it Encyclopedia of Biophysics}, 
edited by G. C. K. Roberts (Springer, Berlin, Heidelberg, 2013), pp. 2149. 

\bibitem{PAH-IR-E} 
C. Joblin, L. d'{H}endecourt, A. L\'{e}ger, and D. D\'{e}fourneau, 
Astron. Astrophys. {\bf 281}, 923 (1994). 

\bibitem{PAH-IR-T} 
C. Joblin, P. Boissel, A. L\'{e}ger, L. d'{H}endecourt, and D. D\'{e}fourneau, 
Astron. Astrophys. {\bf 299}, 835 (1995). 

\bibitem{PAH-IR-R} 
E. Peeters, C. Mackie, A. Candian, and A. G. G. M. Tielens, 
Acc. Chem. Res. {\bf 54}, 1921 (2021). 

\bibitem{Dutta-Chowdhury-2019} 
B. Dutta and J. Chowdhury, 
Chem. Phys. Lett. {\bf 732}, 136645 (2019). 

\bibitem{Ramirez-2004} 
R. Ram\'{\i}rez, T. López-Ciudad, P. Kumar P, and D. Marx, 
J. Chem. Phys. {\bf 121}, 3973 (2004). 

\bibitem{Joalland-2010} 
B. Joalland, M. Rapacioli, A. Simon, C. Joblin, C. J. Marsden, and F. Spiegelman, 
J. Phys. Chem. A {\bf 114}, 5846 (2010). 

\bibitem{Gaigeot-2005} 
M. P. Gaigeot, R. Vuilleumier, M. Sprik, and D. Borgis, 
J. Chem. Theory Comput. {\bf 1}, 772 (2005). 

\bibitem{Gaigeot-2010} 
M.-P. Gaigeot, 
Phys. Chem. Chem. Phys. {\bf 12}, 3336 (2010). 

\bibitem{Vitale-2015} 
V. Vitale, J. Dziedzic, S. M. M. Dubois, H. Fangohr, and C.-K. Skylaris, 
J. Chem. Theory Comput. {\bf 11}, 3321 (2015). 

\bibitem{PAH-IR} 
F. Calvo, M. Basire, and P. Parneix, 
J. Phys. Chem. A {\bf 115}, 8845 (2011). 

\bibitem{matrix-shift} 
J. W. Hastie, R. Hauge, and J. L. Margrave, 
J. Inorg. Nucl. Chem. {\bf 31}, 281 (1969). 

\bibitem{argon-oniom} 
F. Ito, 
Comput. Theor. Chem. {\bf 1161}, 18 (2019). 

\bibitem{WDM1} 
H. R. R\"uter and R. Redmer, 
Phys. Rev. Lett. {\bf 112}, 145007 (2014). 

\bibitem{WDM2} 
V. V. Karasiev, J. W. Dufty, and S. B. Trickey, 
Phys. Rev. Lett. {\bf 120}, 076401 (2018). 

\bibitem{WDM3} 
M. Bonitz \textit{et al.}, 
Phys. Plasmas {\bf 27}, 042710 (2020). 

\end{thebibliography}
\end{document}